\documentclass[preprint,review,12pt,sort&compress]{./template/elsarticle/elsarticle}
\usepackage{epstopdf}
\usepackage{amsmath}
\usepackage{amsfonts}
\usepackage{xcolor}
\usepackage{lineno,hyperref}
\usepackage{placeins} 
\usepackage{array} 
\usepackage{mathtools}
\usepackage{rotating} 
\modulolinenumbers[5]
\pdfoutput=1

\journal{Combustion and Flame}
\newcommand{\sm}[1]{{\scriptscriptstyle#1}} 
\newcolumntype{P}[1]{>{\centering\arraybackslash}p{#1}}
\renewcommand\labelitemi{\quad--}
\newcommand{\tabitem}{~~\llap{\labelitemi}~~}

\begin{document}

\begin{frontmatter}

\title{The Role of Differential Diffusion during Early Flame Kernel Development under Engine Conditions - Part~II: Effect of Flame Structure and Geometry}


\author[mymainaddress]{Tobias Falkenstein}

\author[mymainaddress]{Hongchao Chu}

\author[mymainaddress]{Mathis Bode}

\author[mysecondaryaddress]{Seongwon Kang}

\author[mymainaddress]{Heinz Pitsch\corref{mycorrespondingauthor}}
\cortext[mycorrespondingauthor]{Corresponding author}
\ead{office@itv.rwth-aachen.de}

\address[mymainaddress]{Institute for Combustion Technology, RWTH Aachen University, 52056 Aachen, Germany}
\address[mysecondaryaddress]{Department of Mechanical Engineering, Sogang University, Seoul 121-742, Republic of Korea}

\begin{abstract}
From experimental spark ignition~(SI) engine studies, it is known that the slow-down of early flame kernel development caused by the (${\mathrm{Le}>1}$)-property of common transportation-fuel/air mixtures tends to increase cycle-to-cycle variations~(CCV). To improve the fundamental understanding of the complex phenomena inside the flame structure of developing flame kernels, an engine-relevant DNS database is investigated in this work. Conclusive analyses are enabled by considering equivalent flame kernels and turbulent planar flames computed with ${\mathrm{Le}>1}$ and ${\mathrm{Le}=1}$. 
%
In Part~I of the present study~(Falkenstein et al., Combust. Flame, 2019\nocite{Falkenstein19_kernel_Le_I_cnf}), a reduced representation of the local mixture state was proposed for the purpose of this analysis. Fluctuations in heat release rate attributed to differential diffusion were found to be governed by the parameters local enthalpy, local equivalence ratio, and H-radical mass fraction. Here, a coupling relation for the diffusion-controlled mixture parameter local enthalpy with local flame geometry and structure is derived, characterized by the key parameters $\kappa$ and~${\left|\nabla c \right|/\left|\nabla c \right|_{\mathrm{lam}}}$. 
The analysis shows that the large positive global mean curvature intrinsic to the flame kernel configuration may detrimentally affect the local mixture state inside the reaction zone, 
particularly during the initial flame kernel development phase. External energy supply by spark ignition may effectively bridge over this critical stage, which causes the impact of global mean flame kernel curvature to be small under the present conditions compared to the overall effect of ${\mathrm{Le}\neq1}$ observed in a statistically planar flame. 
Once ignition effects have decayed, the mixture state inside the reaction zone locally exhibits an identical dependence on~$\left|\nabla c \right|$ as in a strained laminar flame. This implies that differential diffusion effects {\color{black}at Karlovitz numbers representative for part-load conditions} are not weakened by small-scale turbulent mixing, which is undesirable for the engine application, but can be favorable in terms of modeling. 
\end{abstract}


\end{frontmatter}

\section{Introduction}
\label{sec:intro}
The reduction of cycle-to-cycle variations~(CCV) in spark ignition~(SI) engines is a prerequisite for the design and control of engines with increased efficiency~\cite{Aleiferis04_lean_ccv,Jung17_lean_si_ccv} and reduced emissions~\cite{Milkins74_CO_HC_ccv_exp,Karvountzis17_emissions_ccv}. From experiments it is well-known that CCV are correlated with the duration of flame kernel development, which was found to be sensitive to stretch effects in ${\mathrm{Le}\neq1}$ mixtures~\cite{Schiffmann17_exp}. The development of models that capture differential diffusion effects during early flame kernel development requires a detailed understanding of the complex parameter interactions inside the unsteady flame structure, which can be well-addressed by analysis of relevant DNS data. 
 {\color{black}To this end, a~DNS database that was designed to be representative for SI~engine part-load conditions is considered in this work. Systematic parameter variations were conducted that facilitate the isolation of effects related to the small flame kernel size respective of the hydrodynamic length scales, as well as of differential diffusion effects. The database consists of five different flame configurations, which have been partly analyzed in previous studies as shown in Tab.~\ref{tab:dns_datasets}. To systematically approach the complexity of early flame kernel development, the overall analysis has been divided into four parts by sequentially considering unity- and non-unity Lewis number flames, as well as macroscopic and micro-scale effects. Run-to-run variations in the global heat release rate of flame kernels  computed in the ${\mathrm{Le}=1}$ limit were attributed to the effect of flame front/turbulence interactions on flame surface area evolution~\cite{Falkenstein19_kernel_Le1_cnf}. In a subsequent study, the investigation of ${\mathrm{Le}=1}$ flames with different $D_{0}/l_{\mathrm{t}}$ (ratio of the initial flame diameter and the integral length scale of turbulence) has shown that such variations in the total flame area are caused by stochastic variations in curvature variance due to large-scale turbulent flow motion with characteristic length scales of at least the flame kernel size~\cite{Falkenstein19_kernel_Le1_jfm}. The present study consists of two parts, where the ${\mathrm{Le}>1}$ datasets are analyzed in detail. 
\begin{table}
\centering
{\color{black}
  \caption{Flame realizations available in the overall DNS database.}
\vspace{0.1cm}
\begin{tabular}{P{.3\textwidth}|P{.12\textwidth}|P{.12\textwidth}|P{.12\textwidth}} 
\hline
~ & ~& \multicolumn{2}{c}{Number of Cases}  \\
Configuration &  $D_{0}/l_{\mathrm{t}}$  & $\mathrm{Le}=1$  & $\mathrm{Le}>1$ \\
~ & ~ & ~ & ~ \\ [-10pt]
\hline
~ & ~ & ~ & ~ \\ [-10pt]
Engine flame kernel & 0.3 & 2 & \textbf{4} \\ 
~ & ~ & ~ & ~ \\ [-10pt]
\hline
~ & ~ & ~ & ~ \\ [-10pt]
Large flame kernel & 2.0 & 1 & ~ \\ 
~ & ~ & ~ & ~ \\ [-10pt]
\hline
~ & ~ & ~ & ~ \\ [-10pt]
Planar flame & $\infty$ & 1 & \textbf{1} \\ 
~ & ~ & ~ & ~ \\ [-10pt]
\hline
\hline
~ & ~ & ~ & ~ \\ [-10pt]
Reference & {} & \cite{Falkenstein19_kernel_Le1_cnf,Falkenstein19_kernel_Le1_jfm} & \cite{Falkenstein19_kernel_Le_I_cnf}, \textbf{this} \\[10pt]
\hline
\end{tabular}
\label{tab:dns_datasets}
}
\end{table} 
In the first part~\cite{Falkenstein19_kernel_Le_I_cnf}, the heat-release-rate response to differential diffusion effects was analyzed based on the integrated chemical source term, which has high practical relevance, and based on the local chemical source term, which was used to identify the governing parameters in ${\mathrm{Le}\neq1}$ flames in a quantitative fashion. In the present second part, the coupling between the local mixture state, which determines the local heat release rate, and the flame geometry and structure is established by pursuing a systematic analytical approach, which will be introduced hereafter.} \par
\begin{sidewaysfigure*}
\centering
\begin{minipage}[b]{0.9\linewidth}
  \graphicspath{{./figures/flame_physics_schematic/inkscape/}}
  \centering
\includegraphics[trim={0cm 0cm 0cm 0cm},clip,width=\linewidth]{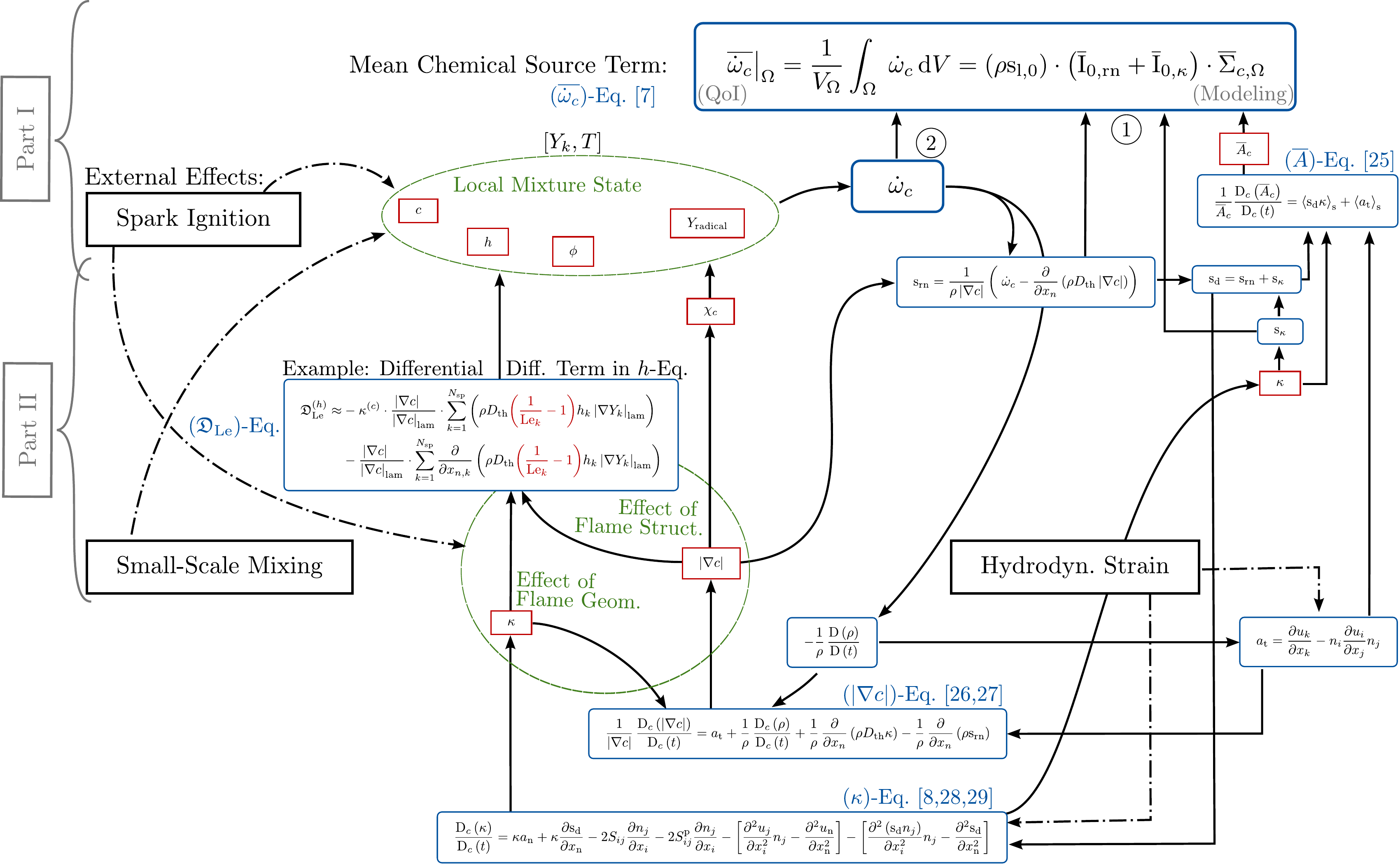}
\nocite{Falkenstein19_kernel_Le1_cnf} 
\nocite{Candel90_fsd_eq,Falkenstein19_kernel_Le1_cnf} 
\nocite{Kim07_sc_align,Sankaran07_sc_grad_eq} 
\nocite{Pope88_surf_turb,Dopazo18_curvEq,Falkenstein19_kernel_Le1_jfm} 
\end{minipage}
\caption{Reduced representation of turbulence/flame interactions on the integrated progress variable source term~$\overline{\dot{\;\omega_{c}}}$ in presence of differential diffusion (${\mathrm{Le}_k\neq1}$). Two analysis pathways are suggested: \textcircled{\tiny 1} A macroscopic perspective based on the FSD concept (cf. r.h.s.\ of \mbox{($\overline{\dot{\;\omega_{c}}}$)-Eq.} in the blue box) considers a propagating flame front with deviations in flame displacement speed from an unstretched laminar flame, similar to laboratory experiments or classical asymptotic theories~\cite{Clavin_Williams_1982,Matalon82_lam_stretch}. \textcircled{\tiny 2} A micro-scale perspective on the local source term~${\dot{\;\omega_{c}}}$, which is determined by the local mixture state, characterized by the local equivalence ratio~$\phi$, enthalpy~$h$, and radical mass fraction~$Y_{\mathrm{radical}}$. These parameters change according to the local flame structure~($\left|\nabla c \right|$) and geometry~($\kappa$), which leads to a response in~${\dot{\;\omega_{c}}}$ to external perturbations of the flame, e.g. by turbulence. {\color{black}In the limit of ${\mathrm{Le}_k=1}$, $\phi$ and~$h$ remain almost constant across the flame structure, i.e.\ the coupling between~${\dot{\;\omega_{c}}}$ and $\left|\nabla c \right|$ and~$\kappa$ is much weaker than in flames with ${\mathrm{Le}_k\neq1}$.} Note that turbulent fluctuations in the scalar fields ahead of the flame have been omitted. For more details, refer to Part~I of the present study~\cite{Falkenstein19_kernel_Le_I_cnf}.}
\label{fig:PrefDiff_schematic}
\end{sidewaysfigure*}

A reduced representation of the governing phenomena in premixed turbulent flames is provided in Fig.~\ref{fig:PrefDiff_schematic} based on~$c$ as a synonym for a reaction progress variable, e.g.\ a (normalized) temperature or a quantity representative for the major product species. We begin the discussion of local effects inside the flame structure that are related to differential diffusion by considering the impact of the local reaction progress variable source term~$\dot{\omega}_c$ (cf.~\textcircled{\tiny 2} in Fig.~\ref{fig:PrefDiff_schematic}), which is related to the local heat release rate. Heat release induces a flame normal propagation velocity~$\mathrm{s}_{\mathrm{rn}}$ and causes flow dilatation, {\color{black}which changes the orientation of the flame front with respect to the flow field,} thus affecting the tangential strain rate $a_{\mathrm{t}}$. These quantities enter the \mbox{($\left|\nabla c \right|$)-Eq.}, which additionally depends on flame curvature~$\kappa$. The local flame structure characterized by $\left|\nabla c \right|$ in turn enters the \mbox{($\kappa$)-Eq.} (through~$\mathrm{s}_{\mathrm{d}}$), which describes the evolution of local flame geometry. Obviously, the turbulent flow field plays a vital role in changing both~$\left|\nabla c \right|$ and~$\kappa$, which is here considered mainly as an external effect on the flame. For a discussion on Lewis-number-dependent feedback from the flame on the surrounding flow field, the reader is referred to the studies by Chakraborty et al.~\cite{Chakraborty11_Le_effect_TKE,Chakraborty16_Le_vort_enstr}. The effect of the local heat release rate on the evolution of flame structure~\cite{Chakraborty08_Le_effect_FSD} and geometry~\cite{Alqallaf19_kernel_curv_eq} was investigated for different Lewis numbers, including the correlation between~$\kappa$ and~$\left|\nabla c \right|$~\cite{Chakraborty05_Le_effect_curv}. Coupling in the opposite direction, i.e.\ \mbox{$\dot{\omega}_c$-response} to changes in flame structure and geometry (cf.\ Fig.~\ref{fig:PrefDiff_schematic}) has been mainly discussed in terms of the competition between focusing and defocusing of heat and reactants~\cite{Rutland93_planar_dns_Le,Klein06_kernel_dns_stretch,Hawkes04_focusing}. 
However, this simplified view based on only two representative scalar fields may not be sufficient to explain all phenomena in multi-species mixtures~\cite{Echekki96_radical_differential_diff,Hilbert04_PRECS_detailed_chem}. 
Recently, Wang et al.~\cite{Wang19_h2_dns_local_equiv_var} conducted a parametric study to assess the effect of nominal equivalence ratio on heat release in high-Karlovitz $\mathrm{H}_2$/air flames. Strong variations in heat release were observed, which correlated with local temperature and local equivalence ratio. The latter two quantities varied significantly across the reaction zone due to differential diffusion effects, particularly under ultra-lean conditions. A reaction pathway analysis revealed the influence of changed radical availability and reaction rate constants on local heat release. It should be noted that at very high Karlovitz numbers, molecular diffusion is overshadowed by small-scale turbulent mixing~\cite{Pitsch00_Le_effect_turb}, which relaxes the local equivalence ratio (and enthalpy) towards the nominal values of the unburned mixture, i.e.\ both quantities may vary less in flames located in the distributed burning regime than in laminar flames~\cite{Aspden11_high_Ka_Le_effects,Savard14_effective_turb_Le}. However, even at high-Karlovitz conditions, differential diffusion was found to be important during turbulent flame development~\cite{Savre13_2d_CH4_devel_flame_high_Ka_phi,Carlsson16_high_Ka_local_phi}. \par
%
%
Under conventional SI engine {\color{black}part-load} conditions, Karlovitz numbers are actually closer to unity and differential diffusion effects substantially reduce the global heat release rate during early flame kernel development, as shown in Part~I of the present study~\cite{Falkenstein19_kernel_Le_I_cnf}. A comparative macroscopic analysis following the \mbox{($\overline{\dot{\;\omega_{c}}}$)-Eq.} in Fig.~\ref{fig:PrefDiff_schematic} attributed the detrimental impact of differential diffusion to the flame normal propagation velocity. A subsequent micro-scale analysis identified the governing parameters that capture the effect of turbulence on the local reaction progress variable source term~${\dot{\;\omega_{c}}}$. Specifically, the local mixture state that determines~${\dot{\;\omega_{c}}}$ was shown to be well-represented by the local equivalence ratio~$\phi$, enthalpy~$h$, and the \mbox{H-radical} mass fraction~$Y_{\mathrm{H}}$. \par 
In the present manuscript, the coupling term between the mixture state parameters and the local flame structure and geometry is exemplarily derived for the enthalpy equation in Sect.~\ref{sec:math_form_h_eq}. The analysis presented in Sect.~\ref{sec:results} demonstrates how the flame-kernel-intrinsic global mean curvature alters the local mixture state and heat release rate. Further,  the effect of turbulence on the flame structure is shown by relating the behavior of flame kernels to strained laminar flames.
\section{Summary of the DNS Database}
\label{sec:dns_database}
Since the DNS database and solution methods for laminar flames have been described in Part~I of the present study~\cite{Falkenstein19_kernel_Le_I_cnf}, only a brief summary of the datasets is provided hereafter. {\color{black}In this work, four different flame configurations (kernel/planar, $\mathrm{Le}=1$/$\mathrm{Le}>1$) are considered (cf.\ Tab.~\ref{tab:dns_datasets}).} The main reference case is a flame kernel developing in homogeneous isotropic turbulence with fully homogeneous, stoichiometric iso-octane/air mixture and ${\mathrm{Le}>1}$. An equivalent dataset has been generated for a statistically planar flame. Additionally, both flame configurations computed in the ${\mathrm{Le}=1}$ limit~\cite{Falkenstein19_kernel_Le1_cnf,Falkenstein19_kernel_Le1_jfm} are available for this study. The flame conditions and non-dimensional groups are listed in Tabs.~\ref{tab:dns_mixture} and~\ref{tab:dns_params}, respectively. 
In Tab.~\ref{tab:dns_mixture}, the mixture thermodynamic state is given by the pressure~$p^{\left(0\right)} $ and temperature of the unburned gas~$T_{\mathrm{u}}$ with equivalence ratio~$\phi_{\mathrm{u}}$. 
\begin{table}
\caption{Flame conditions in the DNS.}
\vspace{0.1cm}
\centering
\begin{tabular}{P{.35\textwidth}|P{.35\textwidth}} 
\hline
~ & ~ \\ [-10pt]
Property & Value  \\
~ & ~ \\ [-10pt]
\hline
~ & ~ \\ [-10pt]
Mixture   & Iso-Octane/Air \\ 
$p^{\left(0\right)} $ & $6$ bar  \\
$T_{\mathrm{u}}$ & $600$ K \\
$\phi_{\mathrm{u}}$ & $1.0$ (homogeneous)\\
$\mathrm{s}_{\mathrm{l}}^{\sm{0}}$ & $\boldsymbol{0.73} \; |\;${\color{black}$0.63$} m/s \\
$l_{\mathrm{f}}$ & $\boldsymbol{69.1}  \; |\;${\color{black}$71.3$} $\mu \mathrm{m}$ \\
Flow Field & Decaying h.i.t. \\
Combust. Regime & Thin Rct. Zones\\
\hline
~ & ~ \\ [-10pt]
$\mathrm{Le}_{\mathrm{eff}}$ & $\boldsymbol{2.0} \; |\; 1.0$ \\
\hline
\end{tabular}
\label{tab:dns_mixture}
\end{table} 
 Note that the turbulent flow field corresponds to decaying homogeneous isotropic turbulence, which explains the DNS parameter ranges given in Tab.~\ref{tab:dns_params}. The turbulent integral length scale and eddy turnover time are denoted by~$l_{\mathrm{t}}$ and~$\tau_{\mathrm{t}}$, $\eta$ is the Kolmogorov length scale, while the laminar flame thickness and initial flame diameter are referred to as~$l_{\mathrm{f}}$ and~$D_{0}$, respectively. \par
\begin{table}
\caption{Engine \citep{Heywood94_COMODIA,Heim11_engine_turb} and DNS characteristic numbers ($\mathrm{Le}>1$).}
\vspace{0.1cm}
\centering
\begin{tabular}{P{.23\textwidth}|P{.23\textwidth}|P{.23\textwidth}}  
\hline
~ & ~ & ~ \\ [-10pt]
Parameter & Engine & DNS ($t_{\mathrm{init}} - t_{\mathrm{end}}$)  \\
~ & ~ & ~ \\ [-10pt]
\hline
~ & ~ & ~ \\ [-10pt]
$\mathrm{Re}_{\mathrm{t}} $ & $100-2390$  & $ 385 - 222$\\[3pt]
$\frac{u_{\mathrm{rms}}}{\mathrm{s}_{\mathrm{l}}^{\sm{0}}}$& $2-15$ & $ 5.9 - 2.8 $\\[3pt]
$\mathrm{Ka} $& $1-6$ & $10.6 - 3.2$\\[3pt]
$\mathrm{Da}$ &$1-100$ & $ 1.9 - 4.6$\\[3pt]
$\frac{l_{\mathrm{t}}}{\eta}$& 100 -200 & $87.2 - 57.2$\\[3pt]  
$\frac{l_{\mathrm{t}}}{l_{\mathrm{f}}}$& $20 - 147$ & $10.9 - 13.0$\\[3pt]
\hline
~ & ~ & ~ \\ [-10pt]
$\frac{D_{0}}{l_{\mathrm{t}}}$ &$<1.0$ & $ 0.3 \; |\; \infty $\\[3pt]
\hline
\end{tabular}
\label{tab:dns_params}
\end{table} 
A brief summary of the DNS setup is given in Tab~\ref{tab:dns_setup}. One important difference between the flame kernel datasets and the planar reference flames is the initialization method. Flame kernels were ignited by a source term in the temperature equation, which results in an early growth phase comparable to engine experiments reported in literature as shown in our previous study~\cite{Falkenstein19_kernel_Le1_jfm}. By contrast, a laminar unstretched flame was imposed into the turbulent flow field as initial condition for the planar-flame DNS to avoid strong dilatation due to the larger burned volume. \par
%
\begin{table}
\caption{Computational setup of the DNS.}
\vspace{0.1cm}
\centering
\begin{tabular}{P{.35\textwidth}|P{.35\textwidth}} 
\hline
~ & ~ \\ [-10pt]
Property & Value  \\
~ & ~ \\ [-10pt]
\hline
~ & ~ \\ [-10pt]
Grid Size   & $960^3$ \\ 
 ~ & ~ \\ [-10pt]
Domain Size & $15 \cdot l_{\mathrm{t}}$ \\ 
 ~ & ~ \\ [-10pt]
Navier-Stokes Eq. & Low-Mach-Approx.~\cite{Mueller98_lowMach} \\
 ~ & ~ \\ [-10pt]
Transport Model & Curt.-H.~\cite{Hirschfelder54_diff_model}, const.-Le  \\
 ~ & ~ \\ [-10pt]
Soret Effect & yes \\
 ~ & ~ \\ [-10pt]
Chem. Mechanism &  26 Spec., based on~\cite{Pitsch96_iso_octane_mech} \\
 ~ & ~ \\ [-10pt]
\multicolumn{1}{l|}{\makebox[20pt][l]{}\color{black}\emph{Flame Kernels:}}  & ~ \\
\multicolumn{1}{l|}{\makebox[20pt][l]{}\tabitem \color{black}Initialization} & \multicolumn{1}{l}{\makebox[10pt][l]{}\tabitem\color{black}Ign. Heat Source}\\
\multicolumn{1}{l|}{\makebox[20pt][l]{}\tabitem \color{black}Boundry. Cond.} & \multicolumn{1}{l}{\makebox[10pt][l]{}\tabitem\color{black}$x,y,z$-dir.: periodic}\\
\multicolumn{1}{l|}{\makebox[20pt][l]{}\tabitem \color{black}Sim. Time} & \multicolumn{1}{l}{\makebox[10pt][l]{}\tabitem\color{black}1x ($t_{\mathrm{sim.}}=3.4\cdot\tau_{\mathrm{t}}$),} \\
\multicolumn{1}{l|}{} & \multicolumn{1}{l}{\makebox[10pt][l]{}\tabitem\color{black}3x ($t_{\mathrm{sim.}}=1.0\cdot\tau_{\mathrm{t}}$)\,} \\
 ~ & ~ \\ [-10pt]
\multicolumn{1}{l|}{\makebox[20pt][l]{}\color{black}\emph{Planar Flame:}}  & ~ \\
\multicolumn{1}{l|}{\makebox[20pt][l]{}\tabitem \color{black}Initialization} & \multicolumn{1}{l}{\makebox[10pt][l]{}\tabitem\color{black}Lam. Flamelet}\\
\multicolumn{1}{l|}{\makebox[20pt][l]{}\tabitem \color{black}Boundry. Cond.} & \multicolumn{1}{l}{\makebox[10pt][l]{}\tabitem\color{black}$x$-dir.: symm./outlet,}\\
\multicolumn{1}{l|}{} & \multicolumn{1}{l}{\makebox[10pt][l]{}\tabitem\color{black}$y,z$-dir.: periodic}\\
\multicolumn{1}{l|}{\makebox[20pt][l]{}\tabitem \color{black}Sim. Time} & \multicolumn{1}{l}{\makebox[10pt][l]{}\tabitem\color{black}($t_{\mathrm{sim.}}=2.8\cdot\tau_{\mathrm{t}}$)} \\
\hline
\end{tabular}
\label{tab:dns_setup}
\end{table} 
\section{Mathematical Formulation: Differential Diffusion Effects in the Enthalpy Equation}
\label{sec:math_form_h_eq}
\begin{figure}
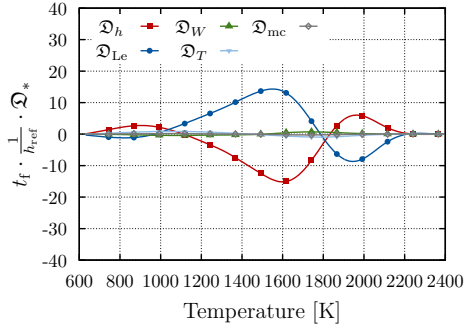

\centering
\begin{minipage}[b]{0.45\textwidth}
  \graphicspath{{./data/FLAME_PLANAR_DNS_01/190409_enth_eq_diff_t_3.6E-04/}}
  \centering
\input{template/change_font_10.tex} 
\scalebox{0.7}{\input{./data/FLAME_PLANAR_DNS_01/190409_enth_eq_diff_t_3.6E-04/planar_dns_CondMean_PDF_Enth_Eq_Terms_vs_T_P01_ltx_JFM_HALF.tex}}
\input{template/change_font_12.tex} 
\end{minipage}
\caption{Conditional mean enthalpy transport terms evaluated in a fully-developed planar turbulent flame with $\mathrm{Le}>1$.}
\label{fig:planar_flame_enth_terms}
\end{figure}

Specifically in stretched flames, differential-diffusion induces variations in element mass fractions (i.e.\ local equivalence ratio)~\cite{deGoey99_stretch_theory,JPope99_local_equiv_ratio_premixed_flame}, enthalpy~\cite{Ashurst87_enth_Le} and radical mass fractions~\cite{Echekki96_radical_differential_diff,Chen00_H2_dns,Aung02_exp_turb_flame_pref_diff} across the flame structure. Consequently, changes in chemical reaction rates and pathways may occur in presence of strong stretch~\cite{Wang19_h2_dns_local_equiv_var}.
In Part~I of the present study~\cite{Falkenstein19_kernel_Le_I_cnf}, 
the combination of ($\phi,h,Y_{\mathrm{H}}$) was identified as the most suitable parameter set to capture differential-diffusion-induced variations in heat release rate conditioned on the reaction progress variable, in agreement with existing literature. Recall that in the present DNS database, the unburned mixture is fully homogeneous, i.e.\ variations in local enthalpy or equivalence ratio develop inside the flame structure (with the exception of the initial spark ignition in case of flame kernels). Hence, the occurrence of inhomogeneous enthalpy and equivalence ratio is mainly a diffusion-driven flame response to external changes in flame structure and geometry. In the following, the mathematical parameter relationship will be derived for the enthalpy in the planar flame dataset. Similarly, one could consider transport equations for the element mass fractions as shown by van~Oijen and de~Goey for a premixed counterflow configuration~\cite{vanOijen02_premixedCntFlow_Le}, which is not pursued here for brevity. While enthalpy was not numerically solved for in the present DNS, a formula consistent with the employed transport model will be derived. We begin with the balance equation for an enthalpy variable $h$ defined as the sum of the sensible and chemical enthalpies~\cite{Poinsot05_book}, here simplified for low-Mach-number flows:
\begin{align}
\frac{\partial \left(\rho h\right)}{\partial t} + \frac{\partial \left(\rho u_j h\right)}{\partial x_j} = &\, \frac{dp^{\left(0\right)}}{dt\;\;\;} 
-\frac{\partial q_j}{\partial x_j}  + \dot{\mathrm{S}}_{\mathrm{ign}} \nonumber \\ 
& + \tau_{ij}\frac{\partial u_i}{\partial x_j} + \dot{q}_{\mathrm{rad}}.
\label{eq:h_all_terms}
\end{align}
In this work, viscous heating and radiative heat losses have been neglected. 
In absence of an ignition heat source {\color{black}and under isobaric conditions, Eq.~(\ref{eq:h_all_terms}) can be simplified as follows}:
\begin{equation}
\begin{aligned}
\frac{\partial \left(\rho h\right)}{\partial t} + \frac{\partial \left(\rho u_j h\right)}{\partial x_j} = & -\frac{\partial q_j}{\partial x_j}.
\label{eq:h_terms_planar}
\end{aligned}
\end{equation}
The enthalpy flux $q_j$ is defined as:
\begin{align}
q_j  = & - \lambda \frac{\partial T}{\partial x_j} + \rho \sum_{k=1}^{N_{\mathrm{sp}}}h_k Y_k V_{j,k} \nonumber \\
 = & - \rho D_{\mathrm{th}} \frac{\partial h}{\partial x_j} + \rho \sum_{k=1}^{N_{\mathrm{sp}}} \left[h_k\left(Y_k V_{j,k} + D_{\mathrm{th}} \frac{\partial Y_k}{\partial x_j}  \right)\right],
\label{eq:h_q_flux}
\end{align}
where the relation ${h=\sum \left(Y_kh_k\right)}$ has been applied. Note that if a simple diffusion model of the form ${V_{j,k} = -\frac{D_{\mathrm{th}}}{\mathrm{Le}_{k}} \frac{1}{Y_k} \frac{\partial Y_k}{\partial x_j}}$ was used and all species Lewis numbers were equal to unity, the last term in Eq.~(\ref{eq:h_q_flux}) would cancel out. The molecular transport expressions employed in this work~\cite{Falkenstein19_kernel_Le1_cnf} yield a more complex net diffusion velocity:
\begin{align}
	V_{j,k} = & \;v_{j,k} + v_j^{\mathrm{mc}} \nonumber \\
 = & -\frac{D_{\mathrm{th}}}{\mathrm{Le}_{k}} \frac{1}{Y_k} \frac{\partial Y_k}{\partial x_j} - \frac{D_{\mathrm{th}}}{\mathrm{Le}_{k}}\frac{1}{W} \frac{\partial W}{\partial x_j} \nonumber \\
 & - \frac{D_{T,k}}{\rho Y_k}\frac{1}{T} \frac{\partial T}{\partial x_j} + \sum_{k=1}^{N_{\mathrm{sp}}} \frac{D_{\mathrm{th}}}{\mathrm{Le}_{n}} \frac{Y_n}{X_n} \frac{\partial X_n}{\partial x_j}.
\label{eq:v_diff_Le}
\end{align}
To obtain a form of the enthalpy equation that explicitly contains the species Lewis numbers, Eqs.~(\ref{eq:v_diff_Le}) and~(\ref{eq:h_q_flux}) are combined as:
\begin{align}
q_j  = & - \rho D_{\mathrm{th}} \frac{\partial h}{\partial x_j} \nonumber \\
& - \rho D_{\mathrm{th}} \sum_{k=1}^{N_{\mathrm{sp}}} \left[\left(\frac{1}{\mathrm{Le}_k} -1 \right)h_k \frac{\partial Y_k}{\partial x_j} \right] \nonumber \\
& - \rho D_{\mathrm{th}} \sum_{k=1}^{N_{\mathrm{sp}}} \left[\frac{1}{\mathrm{Le}_k} Y_k h_k \frac{1}{W} \frac{\partial W}{\partial x_j} \right] \nonumber \\
& + \rho D_{\mathrm{th}} \sum_{k=1}^{N_{\mathrm{sp}}}\left[\frac{1}{\mathrm{Le}_k} Y_k \frac{1}{X_k} \frac{\partial X_k}{\partial x_j} \right] \cdot h \nonumber \\
& -  \sum_{k=1}^{N_{\mathrm{sp}}} \left[D_{T,k} h_k \frac{1}{T} \frac{\partial T}{\partial x_j} \right].
\label{eq:h_q_flux_Le}
\end{align}
Inserting this expression for the enthalpy flux into Eq.~(\ref{eq:h_terms_planar}) yields:
\begin{equation}
\begin{aligned}
\frac{\partial \left(\rho h\right)}{\partial t} + \frac{\partial \left(\rho u_j h\right)}{\partial x_j} = & \underbrace{\frac{\partial }{\partial x_j}\left( \rho D_{\mathrm{th}} \frac{\partial h}{\partial x_j} \right)}_{\displaystyle \mathfrak{D}_{h}} \\
+ & \underbrace{\frac{\partial }{\partial x_j}\left( \rho D_{\mathrm{th}} \sum_{k=1}^{N_{\mathrm{sp}}} \left[\left(\frac{1}{\mathrm{Le}_k} -1 \right)h_k \frac{\partial Y_k}{\partial x_j} \right] \right)}_{\displaystyle \mathfrak{D}_{\mathrm{Le}}} \\
+ & \underbrace{\frac{\partial }{\partial x_j}\left( \rho D_{\mathrm{th}} \sum_{k=1}^{N_{\mathrm{sp}}} \left[\frac{1}{\mathrm{Le}_k} Y_k h_k \frac{1}{W} \frac{\partial W}{\partial x_j} \right]  \right)}_{\displaystyle \mathfrak{D}_{W}} \\
- & \underbrace{\frac{\partial }{\partial x_j}\left( \rho D_{\mathrm{th}} \sum_{k=1}^{N_{\mathrm{sp}}}\left[\frac{1}{\mathrm{Le}_k} Y_k \frac{1}{X_k} \frac{\partial X_k}{\partial x_j} \right] \cdot h \right)}_{\displaystyle \mathfrak{D}_{\mathrm{mc}}} \\
+ & \underbrace{\frac{\partial }{\partial x_j}\left( \sum_{k=1}^{N_{\mathrm{sp}}} \left[D_{T,k} h_k \frac{1}{T} \frac{\partial T}{\partial x_j} \right] \right)}_{\displaystyle \mathfrak{D}_{T}}.
\label{eq:h_terms_planar_Le}
\end{aligned}
\end{equation}
The first term in the r.h.s.\ of Eq.~(\ref{eq:h_terms_planar_Le}) corresponds to molecular diffusion of enthalpy and is denoted as~$\mathfrak{D}_{h}$. Note that~$\mathfrak{D}_{h}$ is independent of nonequal transport properties among different species. Differential diffusion may introduce local variations in enthalpy through~$\mathfrak{D}_{\mathrm{Le}}$, which vanishes in unity-Lewis-number mixtures. As will be shown below, the remaining three terms are comparatively less important in the present flame configurations, but do not cancel out in the limit of mass equidiffusion in an initially gradient-free enthalpy field. The term~$\mathfrak{D}_{W}$ corresponds to diffusion due to spatial variations in molecular weight and~$\mathfrak{D}_{T}$ represents thermodiffusion. Since the applied diffusion model requires a correction term to enforce mass conservation, the associated enthalpy transport appears as~$\mathfrak{D}_{\mathrm{mc}}$ in Eq.~(\ref{eq:h_terms_planar_Le}).
All terms have been evaluated in 
the planar DNS dataset with ${\mathrm{Le}_k\neq 1}$. The results shown in Fig.~\ref{fig:planar_flame_enth_terms} confirm that the differential diffusion term~$\mathfrak{D}_{\mathrm{Le}}$ is of leading order. It is balanced by molecular diffusion of enthalpy through~$\mathfrak{D}_{h}$. 
In order to further investigate the role of differential diffusion in the present turbulent flames, we seek to identify the governing flame parameters which are suitable for conditional averaging of the DNS results and enable a rigorous comparison to laminar reference solutions. \par
After splitting the species mass diffusion terms into components by normal- and tangential diffusion~\cite{Echekki99_diff_term_split}, the differential diffusion term~$\mathfrak{D}_{\mathrm{Le}}$ in Eq.~(\ref{eq:h_terms_planar_Le}) can be rewritten in terms of the scalar-iso-surface curvature $\kappa^{(Y_k)}$ and the scalar-gradient magnitude $\left|\nabla Y_k\right|$:
\begin{equation}
\begin{aligned}
\mathfrak{D}_{\mathrm{Le}}  =&  \sum_{k=1}^{N_{\mathrm{sp}}} \frac{\partial }{\partial x_j}\left( \rho D_{\mathrm{th}} \left(\frac{1}{\mathrm{Le}_k} -1 \right)h_k \frac{\partial Y_k}{\partial x_j} \right) \\
 = & - \sum_{k=1}^{N_{\mathrm{sp}}} \left( \rho D_{\mathrm{th}} \left(\frac{1}{\mathrm{Le}_k} -1 \right)h_k \left|\nabla Y_k\right|  \kappa^{(Y_k)} \right) \\
 & - \sum_{k=1}^{N_{\mathrm{sp}}} \frac{\partial }{\partial x_{n,k}}\left( \rho D_{\mathrm{th}} \left(\frac{1}{\mathrm{Le}_k} -1 \right)h_k \left|\nabla Y_k\right| \right)
\label{eq:h_term_D_Le_kappa_k}
\end{aligned}
\end{equation}
It should be noted that in this work, the curvature~$\kappa$ of iso-surfaces belonging to any scalar field~$\vartheta$ is computed from the normal vector~$n_i$ pointing into the direction of negative scalar gradient, i.e.\ flame kernels have positive global mean curvature:
\begin{align}
\kappa^{(\vartheta)} &= \frac{\partial n_i^{(\vartheta)}}{\partial x_i\;\;}, \label{eq:def_kappa} \\
n_i^{(\vartheta)} &= \frac{-1}{\left|\nabla \vartheta \right|}\frac{\partial \vartheta}{\partial x_i} . \label{eq:def_nvec}
\end{align} 
To parameterize differential diffusion under the  present engine-relevant DNS conditions, the turbulent scalar field shall be related to the laminar unstretched flame in points where $\left|\nabla Y_k\right|_{\mathrm{lam}}>0$. Further, we assume that
\begin{align}
\kappa^{(Y_k)} & \approx \kappa^{(T)} ,  \label{eq:h_term_D_Le_kappa_k_assumpt_T} \\
\frac{\left|\nabla Y_k\right|\quad}{\left|\nabla Y_k\right|_{\mathrm{lam}}} & \approx \frac{\left|\nabla T\right|\quad}{\left|\nabla T\right|_{\mathrm{lam}}}\;, \label{eq:h_term_D_Le_grad_k_assumpt_T}
\end{align}
i.e.\ scalar iso-surfaces are considered to be parallel {\color{black}at a given temperature} and scalar gradients inside the flame structure are proportionally compressed or expanded if the local flame thickness deviates from the unstretched premixed flame. {\color{black}Note that a very similar assumption as the latter was previously used by Lapointe and Blanquart~\cite{Lapointe16_grad_approx_lam} to derive an approximate expression for the turbulent burning velocity (cf.\ Eq.~(A.4) therein).} It should be noted that in presence of strong turbulent micro-mixing, the applied analogy to a laminar reference flame is not expected to hold. Introducing these assumptions into Eq.~(\ref{eq:h_term_D_Le_kappa_k}) yields the approximate differential diffusion term for enthalpy shown in Fig.~\ref{fig:PrefDiff_schematic}:
\begin{equation}
\begin{aligned}
\mathfrak{D}_{\mathrm{Le}}  \approx &  -  \kappa^{(T)} \cdot \frac{\left|\nabla T\right|\quad}{\left|\nabla T\right|_{\mathrm{lam}}}\cdot \sum_{k=1}^{N_{\mathrm{sp}}} \left( \rho D_{\mathrm{th}} \left(\frac{1}{\mathrm{Le}_k} -1 \right)h_k \left|\nabla Y_k\right|_{\mathrm{lam}} \right) \\
 & -  \frac{\left|\nabla T\right|\quad}{\left|\nabla T\right|_{\mathrm{lam}}}\cdot \sum_{k=1}^{N_{\mathrm{sp}}} \frac{\partial }{\partial x_{n,k}}\left( \rho D_{\mathrm{th}} \left(\frac{1}{\mathrm{Le}_k} -1 \right)h_k \left|\nabla Y_k\right|_{\mathrm{lam}} \right)
\label{eq:h_term_D_Le_kappa_k_lam}
\end{aligned}
\end{equation}
As will be shown below, using~$\kappa$ as a measure for local flame geometry and ${\left|\nabla \vartheta \right|/\left|\nabla \vartheta \right|_{\mathrm{lam}}}$ for a representative scalar~$\vartheta$ to characterize changes in flame structure  with respect to a laminar reference flame works reasonably well to parameterize differential diffusion under the present engine-relevant DNS conditions. {\color{black}Obvious choices for~$\vartheta$ include a reaction progress variable or the temperature. Here, the temperature is selected for better comparability with existing literature, since it is commonly used as independent variable in non-unity-Lewis-number flames. However, this choice does not restrict the conclusions to this particular progress variable definition.} \par
{\color{black}It should be mentioned that assumptions~(\ref{eq:h_term_D_Le_kappa_k_assumpt_T}) and~(\ref{eq:h_term_D_Le_grad_k_assumpt_T}) are in fact more restrictive than common flamelet assumptions, which is due to the 
simplicity of the present analytical approach. Here, we seek to link the evolution of the mixture state parameter~$h$ to only one laminar reference flame solution, 
as opposed to a mapping of the full mixture state vector of multiple flamelet solutions to the turbulent flame based on the local flow conditions. While Eq.~(\ref{eq:h_term_D_Le_kappa_k_assumpt_T}) is analogous to neglecting species gradients in $\vartheta$ iso-surface-tangential directions ($\partial/\partial \vartheta_k << \partial/\partial \vartheta$ for $k=2,3$~\cite{Savard17_nonLe_flamelet_model}), the assumption is here introduced for the entire flame structure, i.e.\ not limited to the reaction zone. This is necessary in order to correctly represent differential diffusion effects inside the preheat region, which may significantly affect the enthalpy distribution in the reaction zone. In a suitable flamelet formulation~\cite{Savard17_nonLe_flamelet_model}, the species solution in composition space can adjust to the local curvature of the respective $\vartheta$ iso-surface, and may yield a modeled species field in physical space that violates assumption~(\ref{eq:h_term_D_Le_kappa_k_assumpt_T}). Similarly, the species solution of a laminar flamelet may respond to changes in scalar dissipation rate such that assumption~(\ref{eq:h_term_D_Le_grad_k_assumpt_T}) is not satisfied by a modeled species field. However, it should be noted that the coupling between the local mixture state (e.g.~$h$) and the local flame structure (${\left|\nabla \vartheta \right|/\left|\nabla \vartheta \right|_{\mathrm{lam}}}$) and geometry ($\kappa$) here derived from the separate consideration of molecular diffusion processes in flame  normal and tangential directions in physical space~\cite{Gran96_diff_split} can be analyzed in the same way by considering the enthalpy equation in composition space. Savard and Blanquart~\cite{Savard17_nonLe_flamelet_model} have shown that a premixed flamelet formulation that additionally accounts for the local normal diffusion rate of the representative scalar $\vartheta$ remains valid up to moderate reaction zone Karlovitz numbers of $\mathrm{Ka}_{\delta}<10$, which supports the present assumptions. For a discussion on the differences between a parameterization of the local heat release rate by the diffusion rate, as opposed to~$\left|\nabla \vartheta \right|$ and~$\kappa$, refer to Part~I of the present study~\cite{Falkenstein19_kernel_Le_I_cnf}. Compared to previous work on premixed flamelet equations that contain the parameters~$\left|\nabla \vartheta \right|$~\cite{Savard17_nonLe_flamelet_model} or strain~\cite{Scholtissek19_flamelet}, and~$\kappa$, the present analytical approach pursues a fine-grained, illustrating backward analysis of the governing parameter interactions from the local heat release rate to the external effect of turbulence.} \par
\begin{figure}
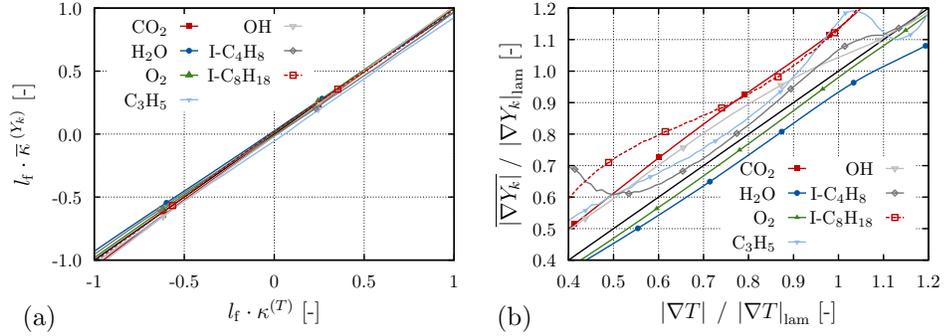

\centering
\begin{minipage}[b]{0.45\textwidth}
  \graphicspath{{./data/FLAME_PLANAR_DNS_01/190407_sc_curv_grad_t_3.6E-04/}}
  \centering
\input{template/change_font_10.tex} 
  \makebox[0pt][l]{\quad(a)}\scalebox{0.7}{\input{./data/FLAME_PLANAR_DNS_01/190407_sc_curv_grad_t_3.6E-04/planar_dns_t3p6e-4_CondMean_jPDF_CurvSpec_vs_CurvT_P01_ltx_JFM_HALF.tex}}
\input{template/change_font_12.tex} 
\end{minipage}
\begin{minipage}[b]{0.45\textwidth}
  \graphicspath{{./data/FLAME_PLANAR_DNS_01/190407_sc_curv_grad_t_3.6E-04/}}
  \centering
\input{template/change_font_10.tex} 
  \makebox[0pt][l]{\quad(b)}\scalebox{0.7}{\input{./data/FLAME_PLANAR_DNS_01/190407_sc_curv_grad_t_3.6E-04/planar_dns_t3p6e-4_CondMean_jPDF_GradSpec_vs_GradT_norm_P01_ltx_JFM_HALF.tex}}
\input{template/change_font_12.tex} 
\end{minipage}
\caption{Correlation of conditional mean curvatures (a) and mean species-gradient ratios (b) with the same quantities evaluated from the temperature field (cf.\ Eqs.~(\ref{eq:h_term_D_Le_kappa_k_assumpt_T}) and~(\ref{eq:h_term_D_Le_grad_k_assumpt_T})). Data has been computed for the fully-developed planar turbulent flame with $\mathrm{Le}>1$.}
\label{fig:planar_flame_pref_diff_param}
\end{figure}

To demonstrate that the assumptions~(\ref{eq:h_term_D_Le_kappa_k_assumpt_T}) and~(\ref{eq:h_term_D_Le_grad_k_assumpt_T}) hold on average under the present conditions, both parameters have been evaluated from different species and based on temperature. Conditional mean results are plotted as function of the temperature-based quantities in Fig.~\ref{fig:planar_flame_pref_diff_param}. Note that the data has been conditioned on non-zero scalar gradients to avoid singularities, but not on specific temperature iso-levels. The strong correlation between flame curvatures computed from different quantities (cf.\ Fig.~\ref{fig:planar_flame_pref_diff_param}\,(a)) confirms the assumption that scalar iso-surfaces in the same spatial location are mostly parallel to each other. Similarly, the correlation between scalar-gradient ratios is shown in Fig.~\ref{fig:planar_flame_pref_diff_param}\,(b). The respective laminar reference gradient is mapped based on the local temperature value. Although the correlation with the temperature-based parameter is less strong compared to the curvature results, the conditional mean gradient ratios evaluated for different species scatter around the diagonal, which is sufficient for the present analysis. {\color{black}Recall that the same laminar unstreteched premixed flame is used as reference flame in all points, i.e.\ the shown correlations are expected to improve if a higher dimensional reference flamelet manifold is used.} \par
In this section, the mathematical relation between the local mixture state (which determines~$\dot{\;\omega_{c}}$) and flame structure and geometry has been derived. Eq.~(\ref{eq:h_term_D_Le_kappa_k_lam}) suggests the following hypothesis for turbulent premixed flames with Karlovitz numbers not much higher than unity: differential diffusion effects approach the limit of a laminar unstretched flame, if the curvature is zero and the local temperature gradient is comparable to the laminar reference flame. This hypothesis and the actual quantitative effects of differential diffusion in both the planar-flame and the engine-relevant flame kernel configuration will be discussed below.
%
\section{Results}
\label{sec:results}
Since differential diffusion introduces a strong coupling between the local mixture state and the flame geometry and structure (cf.\ \mbox{($\mathfrak{D}_{\mathrm{Le}}$)-Eq.} in Fig.~\ref{fig:PrefDiff_schematic}), the particular conditions during early flame kernel development require special attention. In Sect.~\ref{ssec:geom_struct}, differences between the flame kernel and the planar flame configuration that can be attributed to curvature effects will be investigated. The role of turbulence in changing the flame structure will be analyzed in Sect.~\ref{ssec:struct_lam} by relating the behavior of flame kernels to laminar flame solutions. {\color{black}Finally, the role of hydrodynamic strain will be discussed in Sect.~\ref{ssec:strain}}.
\subsection{Combined Effect of Flame Geometry and Structure}
\label{ssec:geom_struct}
In Part~I of this study~\cite{Falkenstein19_kernel_Le_I_cnf} it was shown that the mean heat release rate conditioned on the reaction progress variable (temperature) is significantly reduced by differential diffusion effects in the present flames with $\mathrm{Le}>1$. Further, the local mixture state at a given progress variable was shown to be well-represented by the reduced parameter set ($h, \phi, Y_{\mathrm{H}}$). Here, the difference between the flame kernel configuration that features positive global mean curvature due to the quasi-spherical flame shape and the planar flame configuration will be analyzed. To this end, the evolution of the governing parameters inside the flame structure of flame kernels will be considered in a first step.  \par
\begin{figure}
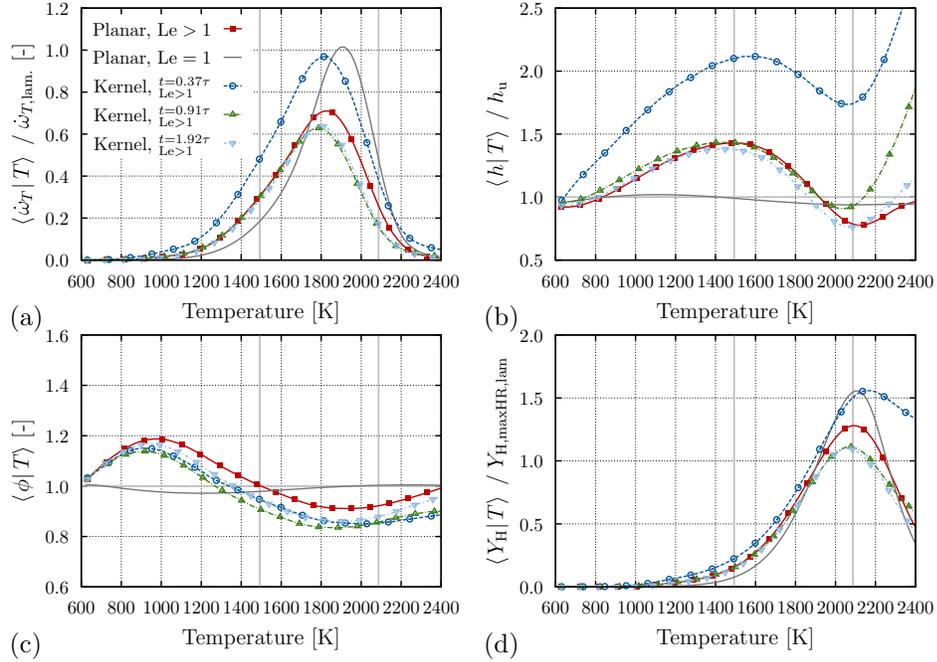
 
\centering
\begin{minipage}[b]{0.45\textwidth}
  \graphicspath{{./data/FLAME_KERNEL_DNS_01/190316_G_local_PHI_curv//}}
  \centering
\input{template/change_font_10.tex} 
  \makebox[0pt][l]{\quad(a)}\scalebox{0.7}{\input{./data/FLAME_KERNEL_DNS_01/190316_G_local_PHI_curv/kernel_dns_CondMean_PDF_HR_vs_T_P01_P02_K01_LOC0123_ltx_JFM_HALF.tex}}
\input{template/change_font_12.tex} 
\end{minipage}
\begin{minipage}[b]{0.45\textwidth}
  \graphicspath{{./data/FLAME_KERNEL_DNS_01/190316_G_local_PHI_curv//}}
  \centering
\input{template/change_font_10.tex} 
  \makebox[0pt][l]{\quad(b)}\scalebox{0.7}{\input{./data/FLAME_KERNEL_DNS_01/190316_G_local_PHI_curv/kernel_dns_CondMean_PDF_Enth_vs_T_P01_P02_K01_LOC0123_ltx_JFM_HALF.tex}}
\input{template/change_font_12.tex} 
\end{minipage}
\begin{minipage}[b]{0.45\textwidth}
  \graphicspath{{./data/FLAME_KERNEL_DNS_01/190316_G_local_PHI_curv//}}
  \centering
\input{template/change_font_10.tex} 
  \makebox[0pt][l]{\quad(c)}\scalebox{0.7}{\input{./data/FLAME_KERNEL_DNS_01/190316_G_local_PHI_curv/kernel_dns_CondMean_PDF_PHI_vs_T_P01_P02_K01_LOC0123_ltx_JFM_HALF.tex}}
\input{template/change_font_12.tex} 
\end{minipage}
\begin{minipage}[b]{0.45\textwidth}
  \graphicspath{{./data/FLAME_KERNEL_DNS_01/190316_G_local_PHI_curv//}}
  \centering
\input{template/change_font_10.tex} 
  \makebox[0pt][l]{\quad(d)}\scalebox{0.7}{\input{./data/FLAME_KERNEL_DNS_01/190316_G_local_PHI_curv/kernel_dns_CondMean_PDF_H_vs_T_P01_P02_K01_LOC0123_ltx_JFM_HALF.tex}}
\input{template/change_font_12.tex} 
\end{minipage}
\caption{Effect of differential diffusion on conditional mean heat release rate~(a) enthalpy~(b), local equivalence ratio~(c), and H-radical mass fraction~(d) as function of temperature for the planar flames and the $\mathrm{Le}>1$ flame kernel configuration. The vertical lines indicate the~$\mathrm{H}_2$-consumption layer~\cite{Pitsch96_iso_octane_mech}. For the flame kernel results, the earlier two times are computed with four realizations, the last time with only one.}
\label{fig:planar_enth_phi_h_vs_temp}
\end{figure}

In Fig.~\ref{fig:planar_enth_phi_h_vs_temp}, the overall impact of differential diffusion is shown by the results extracted from the fully-developed planar flames computed with $\mathrm{Le}>1$ and $\mathrm{Le}=1$. In addition, results for flame kernels with  ${\mathrm{Le}>1}$ are shown. For a spherical expanding flame with $\mathrm{Le}>1$ and substantial mean curvature, one would expect a stronger reduction in heat release rate with respect to the laminar unstretched limit than in case of a statistically planar flame, which is confirmed by the flame kernel results extracted at ${t=1.92\,\tau_{\mathrm{t}}}$ (cf.\ Fig.~\ref{fig:planar_enth_phi_h_vs_temp}\,(a)). At this time instant, compared to the corresponding planar
flame dataset, the mixture state inside the reaction zone (indicated by the two vertical lines) is characterized by reduced levels of both~$h$ and~$\phi$, as well as a lower peak \mbox{H-radical} fraction. During the very early phase of flame kernel development, the detrimental effect of differential diffusion on the heat release rate is effectively compensated by spark ignition, which leads to an increased enthalpy level inside the flame structure (cf.\ Fig.~\ref{fig:planar_enth_phi_h_vs_temp}\,(b)). {\color{black}Additional results for curvature, temperature gradient magnitude, and tangential strain rate across the flame structure that show the effect of high global mean curvature and spark ignition during early flame kernel development are provided in Fig.~\mbox{S-1} of the supplementary material.} \par
\begin{figure}
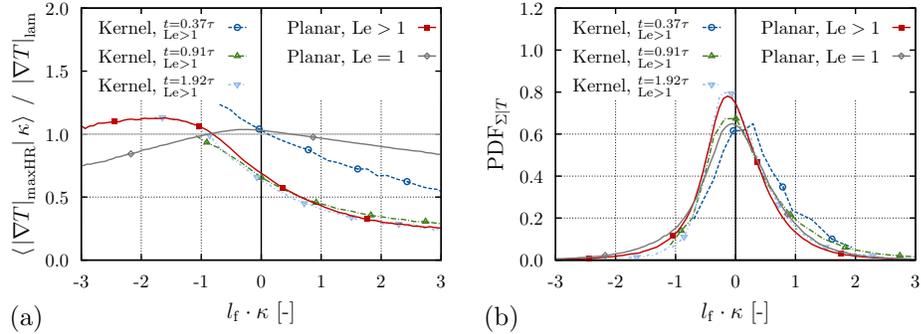
 
\centering
\begin{minipage}[b]{0.45\textwidth}
  \graphicspath{{./data/FLAME_KERNEL_DNS_01/190316_G_local_PHI_curv/}}
  \centering
\input{template/change_font_10.tex} 
\makebox[0pt][l]{\quad(a)}\scalebox{0.7}{\input{./data/FLAME_KERNEL_DNS_01/190316_G_local_PHI_curv/kernel_dns_CondMean_PDF_gradT_CondMaxHR_vs_Curv_K01_LOC0123_P01_P02_allTimes_ltx_JFM_HALF.tex}}
\input{template/change_font_12.tex} 
\end{minipage}
\begin{minipage}[b]{0.45\textwidth}
  \graphicspath{{./data/FLAME_KERNEL_DNS_01/190316_G_local_PHI_curv/}}
  \centering
\input{template/change_font_10.tex} 
\makebox[0pt][l]{\quad(b)}\scalebox{0.7}{\input{./data/FLAME_KERNEL_DNS_01/190316_G_local_PHI_curv/kernel_dns_PDF_vs_Curv_K01_LOC0123_P01_P02_allTimes_ltx_JFM_HALF.tex}}
\input{template/change_font_12.tex} 
\end{minipage}
\caption{Flame Kernel: Normalized conditional mean gradient as function of curvature~(a) {\color{black}and surface-weighted curvature PDF~(b)}. Data has been conditioned on $T_{\mathrm{maxHR}}$. For the flame kernel results, the earlier two times are computed with four realizations, the last time with only one.}
\label{fig:kernel_grad_vs_curv}
\end{figure}

Since mean flame curvature is an intrinsic property of the flame kernel configuration, we will now focus on the flame response to \textit{local curvature}, and use the findings to estimate the effect of \textit{global mean flame curvature} on the heat release rate of flame kernels afterwards. According to Eq.~(\ref{eq:h_term_D_Le_kappa_k_lam}), 
differential-diffusion-induced variations in enthalpy across the flame structure depend on both the local curvature and a representative scalar gradient. Although~$\kappa$ is not explicitly contained in the second term in the r.h.s.\ of Eq.~(\ref{eq:h_term_D_Le_kappa_k_lam}), the occurrence of curvature will typically alter both terms due to the coupling between~$\kappa$ and~$\left|\nabla c \right|$ in premixed flames (cf.\ \mbox{($\kappa$)-} and \mbox{($\left|\nabla c \right|$)-Eqs.} in Fig.~\ref{fig:PrefDiff_schematic}), {\color{black}which can be partly attributed to the negative correlation between the tangential strain rate and curvature~\cite{Kim07_sc_align}, as shown in Fig.~\mbox{S-2} of the supplementary material for the present datasets}. 
To assess the development of the flame structure of the expanding flame kernels after spark ignition, the mean temperature gradient conditioned on curvature has been evaluated for the temperature iso-surface corresponding to the maximum heat release rate in an unstretched premixed flame (`$T_{\mathrm{maxHR}}$'). The correlations shown in Fig.~\ref{fig:kernel_grad_vs_curv}\,(a) are overall very consistent with the ${\mathrm{Le}>1}$ results by Chakraborty and Cant~\cite{Chakraborty05_Le_effect_curv}. For positive curvatures, all flames exhibit a monotonously decreasing trend. Hence, flame kernels that feature mean positive curvature {\color{black}as shown by the curvature PDFs in Fig.~\ref{fig:kernel_grad_vs_curv}\,(b)} will on average have lower temperature gradients compared to the planar flame. With respect to the spark effect it can be concluded that the initially high gradients {\color{black}observed in Fig.~\ref{fig:kernel_grad_vs_curv}\,(a)} decay until ${t=0.91\,\tau_{\mathrm{t}}}$ on the entire $T_{\mathrm{maxHR}}$-iso-surface under the present conditions. It should be noted that in the limit of zero local curvature, the temperature gradient of the turbulent flames is reduced compared to an unstretched laminar flame. This effect will be attributed to hydrodynamic strain later on. Keeping the negative correlation between $\left|\nabla T \right|$ and~$\kappa$ in mind, the combined effect of externally invoked changes in flame geometry~($\kappa$) and structure~($\left|\nabla T \right|$) on the local mixture state ($h, \phi, Y_{\mathrm{H}}$) and heat release rate will be analyzed in the next section. \par
\begin{figure}
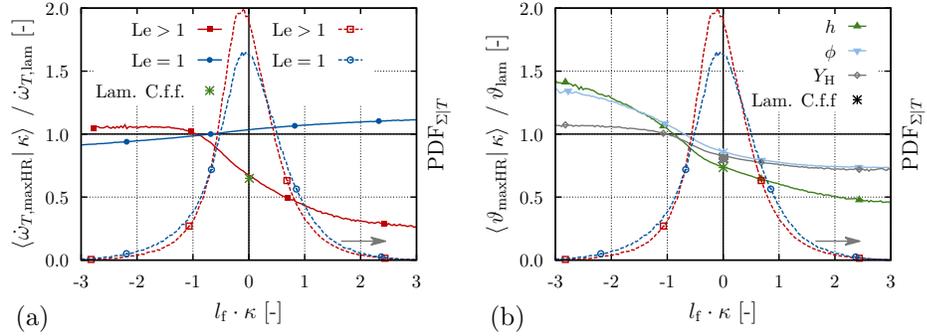
 
\centering
\begin{minipage}[b]{0.45\textwidth}
  \graphicspath{{./data/FLAME_PLANAR_DNS_01/190316_local_equiv_ratio_curv/t_3.600E-04/}}
  \centering
\input{template/change_font_10.tex} 
   \makebox[0pt][l]{\quad(a)}\scalebox{0.7}{\input{./data/FLAME_PLANAR_DNS_01/190316_local_equiv_ratio_curv/t_3.600E-04/planar_dns_t4p8e-4_CondMean_PDF_HR_CondMaxHR_normLam_vs_Curv_P01_P02_ltx_JFM_HALF.tex}}
\input{template/change_font_12.tex} 
\end{minipage}
\begin{minipage}[b]{0.45\textwidth}
  \graphicspath{{./data/FLAME_PLANAR_DNS_01/190316_local_equiv_ratio_curv/t_3.600E-04/}}
  \centering
\input{template/change_font_10.tex} 
    \makebox[0pt][l]{\quad(b)}\scalebox{0.7}{\input{./data/FLAME_PLANAR_DNS_01/190316_local_equiv_ratio_curv/t_3.600E-04/planar_dns_t4p8e-4_CondMean_PDF_Enth_PHI_H_CondMaxHR_vs_Curv_P01_P02_ltx_JFM_HALF.tex}}
\input{template/change_font_12.tex} 
\end{minipage}
\caption{Planar Flame: Conditional mean heat release rate~(a), enthalpy, local equivalence ratio and H-radical mass fraction~(b) as function of curvature. Both panels additionally show the curvature PDF. Data has been conditioned on $T_{\mathrm{maxHR}}$. `C.f.f.'~means counterflow flame.  }
\label{fig:planar_hr_h_phi_h_vs_curv}
\end{figure}

Results for the fully developed planar flames are depicted in Fig.~\ref{fig:planar_hr_h_phi_h_vs_curv}. While the curvature-conditioned mean heat release rates shown in Fig.~\ref{fig:planar_hr_h_phi_h_vs_curv}\,(a) recover similar levels as the corresponding laminar unstretched flames in regions with strong negative curvature (${\kappa < -1/l_{\mathrm{f}}}$), differential diffusion leads to a pronounced negative correlation between curvature and the chemical source term in flame regions with moderately negative as well as positive curvature. This characteristic behavior of ${\mathrm{Le}>1}$ flames has very strong negative consequences for the heat release rate of flame kernels with positive global mean curvature, which will be quantified below. Note that in the planar turbulent flame with ${\mathrm{Le}>1}$, even flame regions with zero curvature exhibit detrimental behavior in terms of the desirable high burning rate for engine applications.
This reduction in heat release rate compared to a laminar unstretched flame is due to the presence of hydrodynamic strain, which is indicated in Fig.~\ref{fig:planar_hr_h_phi_h_vs_curv}\,(a) by the laminar data point extracted from a back-to-back counterflow flame~(c.f.f.) solution. The strained laminar flame solution has been selected based on the condition  ${\left|\nabla T \right|_{{\mathrm{maxHR}},\mathrm{cff}} = \left< \left. \left|\nabla T \right|\; \right|\; \left(T=T_{\mathrm{maxHR}}, \kappa=0\right)\right>}$ and will be analyzed in more detail in Sect.~\ref{ssec:struct_lam} (also refer to \mbox{Fig.~S-3} in the supp.\ material). In both the laminar strained as well as the turbulent flame, the reduction in heat release rate can be explained by the differential-diffusion-induced decrease in enthalpy and equivalence ratio compared to the laminar unstretched limit (cf.\ Fig.~\ref{fig:planar_hr_h_phi_h_vs_curv}\,(b)), i.e.\ by the second term in Eq.~(\ref{eq:h_term_D_Le_kappa_k_lam}). Recall that the transport equations of both quantities do not have a chemical source term (cf.\ Eq.~(\ref{eq:h_terms_planar_Le})), but there is a coupling with the heat release rate through the \mbox{($\left|\nabla T \right|$)-Eq.} as shown in Fig.~\ref{fig:PrefDiff_schematic}. 
%
The  reduction in $Y_{\mathrm{H}}$ shown in Fig.~\ref{fig:planar_hr_h_phi_h_vs_curv}\,(b) is another causal effect for lowering the heat release rate at the considered iso-temperature level~$T_{\mathrm{maxHR}}$. However, the quantitative relation between~${\left.\dot{\omega}_T\right|T}$ and the parameters (${h, \phi, Y_{\mathrm{H}}}$), which were found to well-parametrize the heat release rate across the entire flame structure (cf.\ Part~I~\cite{Falkenstein19_kernel_Le_I_cnf}), shows that the influence of the H-radical mass fraction is much lower than that of~$h$ and~$\phi$ (cf.\ \mbox{Fig.~S-4} of the supp. material). 
With respect to the curvature PDFs plotted in Fig.~\ref{fig:planar_hr_h_phi_h_vs_curv} (${\mathrm{PDF}_{\Sigma|T}}$: surface-weighted PDF, conditioned on ${T=T_{\mathrm{maxHR}}}$), differential diffusion is shown to cause a distribution with shorter tails in the engine-relevant, ${\mathrm{Le}>1}$ case, consistent with the analysis by Alqallaf et al.~\cite{Alqallaf19_kernel_curv_eq}. 
Between the $10^{\mathrm{th}}$ and $90^{\mathrm{th}}$ percentile of the ${\mathrm{Le}>1}$ curvature distribution, the normalized enthalpy and local equivalence ratio vary in the ranges $h/h_{\mathrm{lam}}=0.97...0.63$ and $\phi=1.05...0.81$, which are strong deviations from the ${\mathrm{Le}=1}$ limit with $h/h_{\mathrm{lam}}\approx 1$ and $\phi\approx 1$. 
Since the key difference between the considered planar flames and the engine-relevant flame kernel configuration is the curvature distribution {\color{black}(cf. Fig.~\ref{fig:kernel_grad_vs_curv})}, the effect on the mean heat release rate will be estimated in the following. \par
\begin{figure}
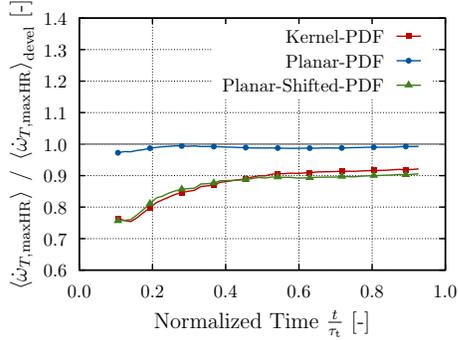
 
\centering
\begin{minipage}[b]{0.45\textwidth}
  \graphicspath{{./data/FLAME_KERNEL_DNS_01/190617_curvPDF_effect_HR/}}
  \centering
\input{template/change_font_10.tex} 
\scalebox{0.7}{\input{./data/FLAME_KERNEL_DNS_01/190617_curvPDF_effect_HR/curvPDF_effect_HR_time_LOC01_P01_JFM_HALF.tex}}
\input{template/change_font_12.tex} 
\end{minipage}
\caption{Convolution of different time-varying curvature distributions with the conditional mean heat release rate of a fully developed turbulent planar flame with ${\mathrm{Le}>1}$ (cf.\ Fig.~\ref{fig:planar_hr_h_phi_h_vs_curv}\,(a)). Data has been conditioned on $T_{\mathrm{maxHR}}$.}
\label{fig:hr_curv_pdf_vs_time}
\end{figure}

The curvature distribution of flame kernels exhibits positive mean curvature, which is expected to reduce the mean heat release rate due to differential diffusion effects (cf.\ Fig.~\ref{fig:planar_hr_h_phi_h_vs_curv}\,(a)). Further, the very early phase of flame kernel development is characterized by the occurrence of skewness towards high positive curvature, which has been previously shown in the ${\mathrm{Le}=1}$ limit~\cite{Falkenstein19_kernel_Le1_jfm} and is confirmed in \mbox{Fig.~S-5} of the supplementary material for the present ${\mathrm{Le}>1}$ datasets. 
To investigate the impact of the curvature distribution, we have convoluted the curvature-conditioned mean heat release rate of the fully developed planar flame that was shown in Fig.~\ref{fig:planar_hr_h_phi_h_vs_curv}\,(a) with three different curvature distributions: i) the actual curvature PDF of the ${\mathrm{Le}>1}$ flame kernel configuration (four realizations), ii) the time-varying curvature PDF of the planar flame (to assess the effect of the kernel's mean positive curvature), and iii) an artificial curvature PDF generated from the PDF of the planar flame, but shifted to the same mean curvature as the flame kernel and renormalized (to test the effect of positve curvature skewness). This procedure has been repeated for a sequence of time instances during early flame development, while keeping the conditional mean heat release rate constant. Hence, the effect of \textit{mean curvature} intrinsic to the flame kernel configuration on the mean heat release rate can be separated from differences in \textit{curvature PDF shape} and spark effects. Additionally, this approach may be a first estimate for the performance of an ideal model {\color{black}that approximates the global heat release rate of flame kernels based on the mean flame kernel curvature and statistics of a fully developed planar flame, which is further investigated in~\ref{apx:model_assess}. The mean heat release rates resulting from the convolution on the $T_{\mathrm{maxHR}}$-iso-surface} are shown in Fig.~\ref{fig:hr_curv_pdf_vs_time} as function of time. 
According to this estimate, mean curvature may have a significant effect on the mean heat release rate during the very early phase of flame kernel development. 
However, it was shown in Fig.~\ref{fig:planar_enth_phi_h_vs_temp}\,(a) that spark ignition initially leads to a higher heat release rate than in the planar flame, which is not accounted for in Fig.~\ref{fig:hr_curv_pdf_vs_time}. This means that the initially strong detrimental effect of differential diffusion induced by high positive mean curvature of flame kernels (cf.\ Fig.~\ref{fig:hr_curv_pdf_vs_time}) is effectively compensated by external enthalpy supply under the present conditions. After a time of ${t\approx 0.5\,\tau_{\mathrm{t}}}$ is reached, the reduction compared to the developing planar flame amounts to approximately 10\,\%. Since the same applies for the planar-shifted PDF convolution, this reduction is only the effect of mean curvature of the kernel and the positive curvature skewness has only a minor effect. Also note that after ${t\approx 0.5\,\tau_{\mathrm{t}}}$, the relative difference in mean heat release rate between the flame kernel and the planar flame estimated in Fig.~\ref{fig:hr_curv_pdf_vs_time} and observed at $T_{\mathrm{maxHR}}$ in Fig.~\ref{fig:planar_enth_phi_h_vs_temp}\,(a) are very similar, which supports the validity of the preceding analysis. \par
Hence, it can be concluded that compared to the overall effect of differential diffusion reflected by the difference between the ${\mathrm{Le}>1}$/${\mathrm{Le}=1}$ planar-flame results in Fig.~\ref{fig:planar_enth_phi_h_vs_temp}\,(a), the additional reduction in mean heat release rate due to to positive mean curvature of flame kernels is rather small, thanks to initial ignition energy supply. 
%
It should also be noted that the minor influence of the PDF-shape on the mean heat release rate does not imply that characteristic kernel/turbulence interactions are per se irrelevant for flame kernel development. In fact, the skewness of the curvature distribution might be an indicator for extinction tendency in the broken kernel regime~\cite{Echekki07_kernel_vortex_map_01,Vasudeo10_kernel_vortex_map_01,Reddy11_regime_map}, which is a topic suggested for future work. \par
\subsection{Dependence of ($h, \phi, Y_{\mathrm{H}}$) on Local Variations in Flame Structure for a given Curvature}
\label{ssec:struct_lam}
According to Fig.~\ref{fig:PrefDiff_schematic}, the local heat release rate is determined by the local mixture state, characterized by the parameter set (${c, h, \phi, Y_{\mathrm{H}}}$). For ${\mathrm{Le}_k\ne1}$, $h$ has been shown to be coupled to the local flame structure~($\left|\nabla c \right|$) and geometry~($\kappa$) through the \mbox{($\mathfrak{D}_{\mathrm{Le}}$)-term} in the enthalpy transport equation (cf.\ Eq.~(\ref{eq:h_term_D_Le_kappa_k_lam})). While the role of flame curvature has been already discussed in Sect.~\ref{ssec:geom_struct}, the objective here is to quantify the effect of externally invoked changes in~$\left|\nabla c \right|$ due to hydrodynamic strain (cf.\ \mbox{($\left|\nabla c \right|$)-Eq.} in Fig.~\ref{fig:PrefDiff_schematic}). In the limit of ${\kappa=0}$, this analysis can be consistently applied to turbulent flames and the laminar counterflow flame~(c.f.f.) configuration, which has been already used as a reference solution in Fig.~\ref{fig:planar_hr_h_phi_h_vs_curv}. In the following sections, we will first show how the planar flame compares to the laminar counterflow configuration and relate flame kernel development to the fully developed planar results afterwards. For clarity, the analysis will be limited to the $T_{\mathrm{maxHR}}$-iso-surfaces. \par
\begin{figure}
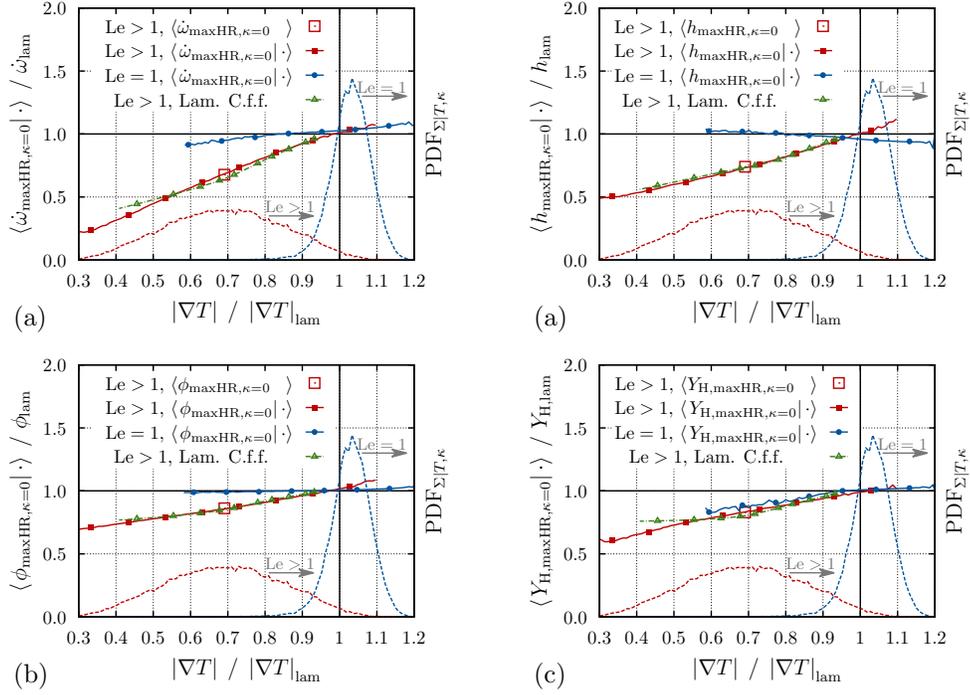
 
\centering
\begin{minipage}[b]{0.45\textwidth}
  \graphicspath{{./data/FLAME_PLANAR_DNS_01/190316_local_equiv_ratio_curv/t_3.600E-04/}}
  \centering
\input{template/change_font_10.tex} 
  \makebox[0pt][l]{\quad(a)}\scalebox{0.7}{\input{./data/FLAME_PLANAR_DNS_01/190316_local_equiv_ratio_curv/t_3.600E-04/planar_dns_t4p8e-4_CondMean_PDF_HR_CondMaxHR_vs_gradT_P01_ltx_JFM_HALF.tex}}
\input{template/change_font_12.tex} 
\end{minipage}
\hfill\vspace{0.2cm}
\begin{minipage}[b]{0.45\textwidth}
  \graphicspath{{./data/FLAME_PLANAR_DNS_01/190316_local_equiv_ratio_curv/t_3.600E-04/}}
  \centering
\input{template/change_font_10.tex} 
  \makebox[0pt][l]{\quad(a)}\scalebox{0.7}{\input{./data/FLAME_PLANAR_DNS_01/190316_local_equiv_ratio_curv/t_3.600E-04/planar_dns_t4p8e-4_CondMean_PDF_Enth_CondMaxHR_vs_gradT_P01_ltx_JFM_HALF.tex}}
\input{template/change_font_12.tex} 
\end{minipage}
\hfill\vspace{0.2cm}
\begin{minipage}[b]{0.45\textwidth}
  \graphicspath{{./data/FLAME_PLANAR_DNS_01/190316_local_equiv_ratio_curv/t_3.600E-04/}}
  \centering
\input{template/change_font_10.tex} 
    \makebox[0pt][l]{\quad(b)}\scalebox{0.7}{\input{./data/FLAME_PLANAR_DNS_01/190316_local_equiv_ratio_curv/t_3.600E-04/planar_dns_t4p8e-4_CondMean_PDF_phi_CondMaxHR_vs_gradT_P01_ltx_JFM_HALF.tex}}
\input{template/change_font_12.tex} 
\end{minipage}
\hfill\vspace{0.1cm}
\begin{minipage}[b]{0.45\textwidth}
  \graphicspath{{./data/FLAME_PLANAR_DNS_01/190316_local_equiv_ratio_curv/t_3.600E-04/}}
  \centering
\input{template/change_font_10.tex} 
  \makebox[0pt][l]{\quad(c)}\scalebox{0.7}{\input{./data/FLAME_PLANAR_DNS_01/190316_local_equiv_ratio_curv/t_3.600E-04/planar_dns_t4p8e-4_CondMean_PDF_H_CondMaxHR_vs_gradT_P01_ltx_JFM_HALF.tex}}
\input{template/change_font_12.tex} 
\end{minipage}
\hfill\vspace{0.1cm}
\caption{Planar Flame: Conditional mean heat release rate~(a), enthalpy~(b), local equivalence ratio~(c), and H-radical mass fraction~(d) as function of normalized gradient. Data has been conditioned on $T_{\mathrm{maxHR}}$ and $\kappa=0$. }
\label{fig:planar_h_phi_vs_grad}
\end{figure}

In Fig.~\ref{fig:planar_h_phi_vs_grad}, the flame structure of both turbulent and laminar flames is parameterized by~$\left|\nabla T \right|$, which is analogous to employing the scalar dissipation rate~$\chi_T$ for modeling purposes~\cite{Kolla10_strained_flamelet_model}. It is shown that in the fully developed planar flame with ${\mathrm{Le}>1}$, local regions with zero curvature on average exhibit~$\dot{\omega}_T$ and ($h, \phi, Y_{\mathrm{H}}$) dependencies on~$\left|\nabla T\right|$ that are almost identical to the laminar premixed counterflow configuration. If differential diffusion effects were significantly reduced by small-scale turbulent mixing, the slope of the conditional means would approach the plotted ${\mathrm{Le}=1}$ results~\cite{Savard14_effective_turb_Le,Savard15_Le_effects_C7H16_highKa,Savard17_nonLe_flamelet_model}. This is an important finding since the Karlovitz numbers in the present datasets are located in the upper range of conventional SI engine {\color{black}part-load} conditions. Note that the behavior of an unstretched laminar flame is locally recovered for $\left|\nabla T \right|=\left|\nabla T \right|_{\mathrm{lam}}$, which confirms the chosen parametrization derived from the \mbox{($\mathfrak{D}_{\mathrm{Le}}$)-Eq.} The unconditional mean value plotted as open square symbol at ${\,\left|\nabla T \right|/\left|\nabla T \right|_{\mathrm{lam}}\approx 0.7}$ in each subfigure marks the unfavorable conditions found on average near $T_{\mathrm{maxHR}}$-iso-surface elements with $\kappa = 0$ (also cf.\ Fig.~\ref{fig:planar_hr_h_phi_h_vs_curv}). As a remark to the shown correlations for a constant iso-temperature, it should be mentioned that each quantity may respond to strain by changes in the overall magnitude, or by qualitatively changing its distribution across the flame structure towards lower/higher temperatures (cf.\ \mbox{Fig.~S-3} in the supp.\ material). \par
Regarding the conditional \mbox{$\left|\nabla T \right|$-PDFs} ($\mathrm{PDF}_{\Sigma|T,\,\kappa}$: surface-weighted PDF, conditioned on ${T=T_{\mathrm{maxHR}}}$ and ${\kappa=0}$) shown in Fig.~\ref{fig:planar_h_phi_vs_grad}, large variance can be observed in the ${\mathrm{Le}>1}$ flame, besides the significantly reduced mean value compared to the ${\mathrm{Le}=1}$ dataset. The large variance can be explained by the strong coupling between flame structure and geometry with the local mixture state by differential diffusion as sketched in Fig.~\ref{fig:PrefDiff_schematic}. Any change in scalar gradients may trigger changes in enthalpy and local equivalence ratio (cf.\ \mbox{($\mathfrak{D}_{\mathrm{Le}}$)-Eq.}), which cause direct response in heat release rate. Since scalar gradient production depends on dilatation through the tangential strain term (cf.\ \mbox{($\left|\nabla c \right|$)-Eq.}), a wider gradient distribution can be expected in ${\mathrm{Le}\neq1}$ flames. In the ${\mathrm{Le}=1}$ limit, external changes in~$\left|\nabla T \right|$ or~$\kappa$ trigger only minor variations in the local mixture state due to the Soret effect since the \mbox{($\mathfrak{D}_{\mathrm{Le}}$)-Eq.} vanishes. \par
\begin{figure}
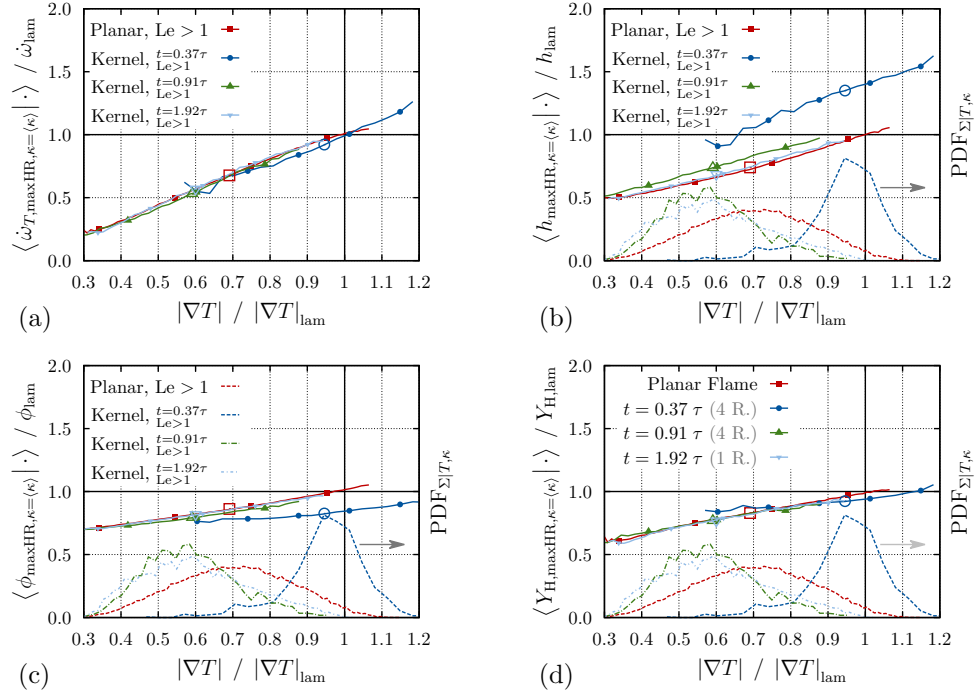
 
\centering
\begin{minipage}[b]{0.45\textwidth}
  \graphicspath{{./data/FLAME_KERNEL_DNS_01/190316_G_local_PHI_curv/}}
  \centering
\input{template/change_font_10.tex} 
  \makebox[0pt][l]{\quad(a)}\scalebox{0.7}{\input{./data/FLAME_KERNEL_DNS_01/190316_G_local_PHI_curv/kernel_dns_CondMean_PDF_HR_CondMaxHR_vs_gradT_curv_uncondMean_K01_LOC0123_P01_ltx_JFM_HALF_noPDF.tex}}
\input{template/change_font_12.tex} 
\end{minipage}
\hfill\vspace{0.2cm}
\begin{minipage}[b]{0.45\textwidth}
  \graphicspath{{./data/FLAME_KERNEL_DNS_01/190316_G_local_PHI_curv/}}
  \centering
\input{template/change_font_10.tex} 
  \makebox[0pt][l]{\quad(b)}\scalebox{0.7}{\input{./data/FLAME_KERNEL_DNS_01/190316_G_local_PHI_curv/kernel_dns_CondMean_PDF_Enth_CondMaxHR_vs_gradT_curv_uncondMean_K01_LOC0123_P01_ltx_JFM_HALF.tex}}
\input{template/change_font_12.tex} 
\end{minipage}
\hfill\vspace{0.2cm}
\begin{minipage}[b]{0.45\textwidth}
  \graphicspath{{./data/FLAME_KERNEL_DNS_01/190316_G_local_PHI_curv/}}
  \centering
\input{template/change_font_10.tex} 
    \makebox[0pt][l]{\quad(c)}\scalebox{0.7}{\input{./data/FLAME_KERNEL_DNS_01/190316_G_local_PHI_curv/kernel_dns_CondMean_PDF_phi_CondMaxHR_vs_gradT_curv_uncondMean_K01_LOC0123_P01_ltx_JFM_HALF.tex}}
\input{template/change_font_12.tex} 
\end{minipage}
\hfill\vspace{0.1cm}
\begin{minipage}[b]{0.45\textwidth}
  \graphicspath{{./data/FLAME_KERNEL_DNS_01/190316_G_local_PHI_curv/}}
  \centering
\input{template/change_font_10.tex} 
  \makebox[0pt][l]{\quad(d)}\scalebox{0.7}{\input{./data/FLAME_KERNEL_DNS_01/190316_G_local_PHI_curv/kernel_dns_CondMean_PDF_H_CondMaxHR_vs_gradT_curv_uncondMean_K01_LOC0123_P01_ltx_JFM_HALF.tex}}
\input{template/change_font_12.tex} 
\end{minipage}
\hfill\vspace{0.1cm}
\caption{{\color{black}Conditional mean heat release rate~(a)}, enthalpy (b), local equivalence ratio (c), and {\color{black}H-radical mass fraction~(d)} as function of normalized gradient. Data has been conditioned on $T_{\mathrm{maxHR}}$ and $\kappa=\left<\kappa \right>$. The open symbols denote~$\left<\dot{\omega}_{T,\mathrm{maxHR},\kappa =\left<\kappa \right> }\right>$, $\left<h_{\mathrm{maxHR},\kappa =\left<\kappa \right> }\right>$, $\left<\phi_{\mathrm{maxHR},\kappa =\left<\kappa \right> }\right>$ and~$\left<Y_{\mathrm{H,maxHR},\kappa =\left<\kappa \right> }\right>$, respectively. For the flame kernel results, the earlier two times are computed with four realizations, the last time with only one.}
\label{fig:kernel_h_phi_vs_grad}
\end{figure}

Now that relationships between the canonical configuration of a fully developed turbulent planar flame and strained laminar flame characteristics have been established, we seek to connect early flame kernel development to the planar-flame results. 
Recall that in contrast to the afore-discussed planar flames, flame kernels were ignited with a volumetric heat source of spherical shape, which provides excess enthalpy (cf.\ Fig.~\ref{fig:planar_enth_phi_h_vs_temp}\,(b)) and high scalar gradients (cf.\ Fig.~\ref{fig:kernel_grad_vs_curv}). 
In Fig.~\ref{fig:kernel_h_phi_vs_grad}, {\color{black}the mean heat release rate and mixture state parameters} conditioned on~$\left|\nabla T \right|$ are shown for three different time instances of flame kernel development. As before, the data has been conditioned on ($T_{\mathrm{maxHR}}$)-iso-surface elements. Here, the data has been conditioned on the respective flame-averaged curvature~${\left<\kappa\right>}$, i.e.\ for the planar flame the same data as in Fig.~\ref{fig:planar_h_phi_vs_grad} is shown due to ${\left<\kappa\right>=0}$.  {\color{black}First, we focus on the development of the flame structure in time. The results in Fig.~\ref{fig:kernel_h_phi_vs_grad} suggest that on average, the dependence of the heat release rate and the H-radical mass fraction on~$\left|\nabla T \right|$ barely changes over time. This can be explained by the coupling between the heat release rate and the evolution of scalar gradients (cf.\ \mbox{($\left|\nabla c \right|$)-Eq.} in Fig.~\ref{fig:PrefDiff_schematic}), and by the fact that $Y_{\mathrm{H}}$ contains the chemical response to changes in the local mixture state. By contrast, the diffusion-controlled parameters enthalpy and local equivalence ratio that do not have a chemical source term in their transport equation exhibit strong deviations during the early kernel development phase. Spark ignition initially supplies the flame structure with excess enthalpy leading to high heat release (cf. open circle in Fig.~\ref{fig:kernel_h_phi_vs_grad}\,(a)) and produces large gradients, as indicated by the \mbox{$\left|\nabla T \right|$-PDF} at ${t=0.37\,\tau_{\mathrm{t}}}$. The initial deviation of the local equivalence ratio from the planar flame results (cf.\ Fig.~\ref{fig:kernel_h_phi_vs_grad}\,(c)) is likely due to the development of the characteristic species distribution inside the highly curved flame structure in presence of large temperature gradients generated by spark ignition.}
At ${t=0.91\,\tau_{\mathrm{t}}}$, the correlations with~$\left|\nabla T \right|$ are overall fairly similar to the developed planar flame. However, the gradient distribution indicates generally reduced gradients, which is due to the presence of mean flame curvature (cf.\ Fig.~\ref{fig:kernel_grad_vs_curv}). At ${t=1.92\,\tau_{\mathrm{t}}}$, a tendency towards the gradient distribution of the developed flame can be observed, but still with significantly higher probability of low gradients. In summary, the ${\mathrm{Le}>1}$ flame kernels approach very similar conditional dependencies as the planar turbulent and strained laminar flames, after the initial non-equilibrium correlation between~$\kappa$ and $\left|\nabla T\right|$ imposed by spark ignition has decayed within less than two laminar flame times. Differences between the iso-surface averaged quantities in the flame kernel and the planar flame configuration can be attributed to the overall larger flame thickness caused by the positive mean flame kernel curvature. \par
%
\subsection{The Role of Hydrodynamic Strain in the Limit of Zero Curvature}
\label{ssec:strain}
\begin{figure}
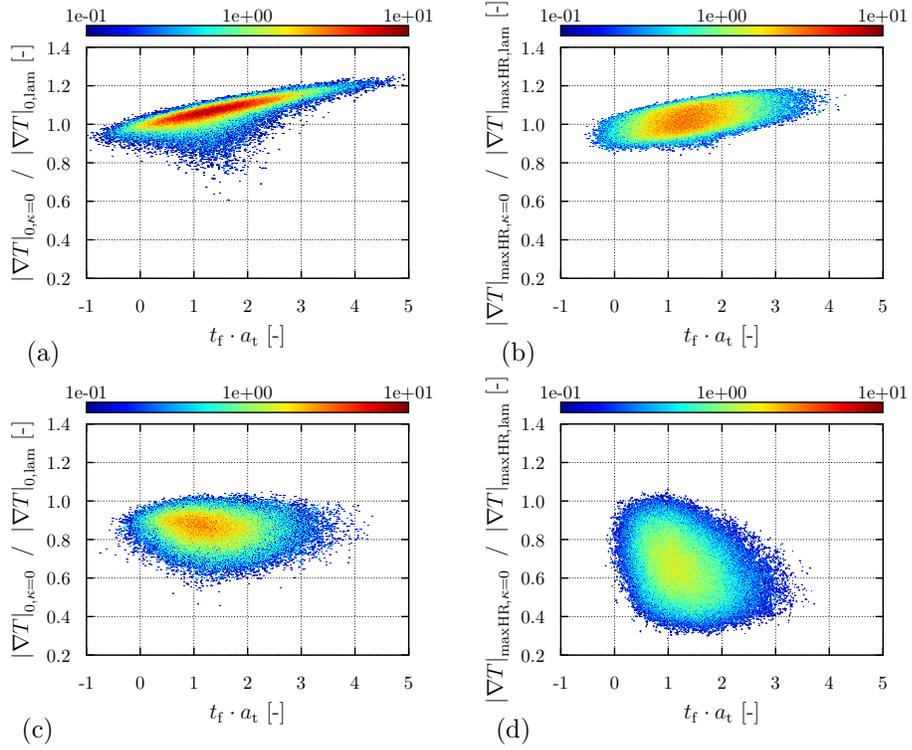
 
\centering
\begin{minipage}[b]{0.45\textwidth}
  \graphicspath{{./data/FLAME_PLANAR_DNS_02/200319_strain_effect_all//}}
  \centering
\input{template/change_font_10.tex} 
\scalebox{0.7}{\input{./data/FLAME_PLANAR_DNS_02/200319_strain_effect_all/planar_dns_jPDF_gradT_vs_strain_curv0_T0_P02_ltx_JFM_HALF.tex}}
\input{template/change_font_12.tex} 
\end{minipage}
\vspace{-0.2cm}
\begin{minipage}[b]{0.45\textwidth}
  \graphicspath{{./data/FLAME_PLANAR_DNS_02/200319_strain_effect_all//}}
  \centering
\input{template/change_font_10.tex} 
\scalebox{0.7}{\input{./data/FLAME_PLANAR_DNS_02/200319_strain_effect_all/planar_dns_jPDF_gradT_vs_strain_curv0_TmaxHR_P02_ltx_JFM_HALF.tex}}
\input{template/change_font_12.tex} 
\end{minipage}
\vspace{-0.2cm}
\begin{minipage}[b]{0.45\textwidth}
\input{template/change_font_10.tex} 
\quad(a)
\input{template/change_font_12.tex} 
\end{minipage}
\vspace{0.3cm}
\begin{minipage}[b]{0.45\textwidth}
\input{template/change_font_10.tex} 
\quad(b)
\input{template/change_font_12.tex} 
\end{minipage}
\vspace{0.3cm}
\begin{minipage}[b]{0.45\textwidth}
  \graphicspath{{./data/FLAME_PLANAR_DNS_01/200319_strain_effect_all//}}
  \centering
\input{template/change_font_10.tex} 
\scalebox{0.7}{\input{./data/FLAME_PLANAR_DNS_01/200319_strain_effect_all/planar_dns_jPDF_gradT_vs_strain_curv0_T0_P01_ltx_JFM_HALF.tex}}
\input{template/change_font_12.tex} 
\end{minipage}
\vspace{-0.5cm}
\begin{minipage}[b]{0.45\textwidth}
  \graphicspath{{./data/FLAME_PLANAR_DNS_01/200319_strain_effect_all//}}
  \centering
\input{template/change_font_10.tex} 
\scalebox{0.7}{\input{./data/FLAME_PLANAR_DNS_01/200319_strain_effect_all/planar_dns_jPDF_gradT_vs_strain_curv0_TmaxHR_P01_ltx_JFM_HALF.tex}}
\input{template/change_font_12.tex} 
\end{minipage}
\vspace{-0.5cm}
\begin{minipage}[b]{0.45\textwidth}
\input{template/change_font_10.tex} 
\quad(c)
\input{template/change_font_12.tex} 
\end{minipage}
\vspace{0.2cm}
\begin{minipage}[b]{0.45\textwidth}
\input{template/change_font_10.tex} 
\quad(d)
\input{template/change_font_12.tex} 
\end{minipage}
\vspace{0.2cm}
\caption{{\color{black}Planar Flame: Joint-PDFs of the temperature gradient magnitude and the tangential strain rate for the ${\mathrm{Le}=1}$ dataset~(a,b) and the engine-relevant ${\mathrm{Le}>1}$ flame~(c,d). The iso-temperatures $T_0$ and $T_{\mathrm{maxHR}}$ correspond to the maximum temperature gradient and maximum heat release rate in the corresponding laminar unstretched premixed flames, respectively. The data has been conditioned on $\kappa=0$.} }
\label{fig:planar_jPDF_grad_vs_strain}
\end{figure}

\begin{figure}
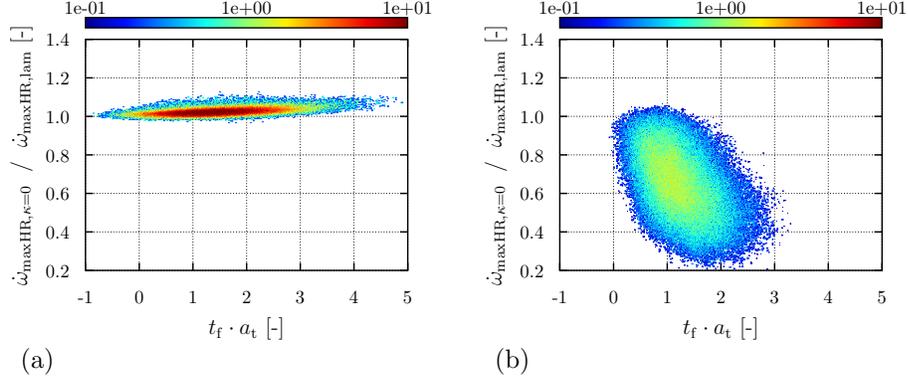
 
\centering
\begin{minipage}[b]{0.45\textwidth}
  \graphicspath{{./data/FLAME_PLANAR_DNS_02/200319_strain_effect_all//}}
  \centering
\input{template/change_font_10.tex} 
\scalebox{0.7}{\input{./data/FLAME_PLANAR_DNS_02/200319_strain_effect_all/planar_dns_jPDF_HR_vs_strain_curv0_TmaxHR_P02_ltx_JFM_HALF.tex}}
\input{template/change_font_12.tex} 
\end{minipage}
\begin{minipage}[b]{0.45\textwidth}
  \graphicspath{{./data/FLAME_PLANAR_DNS_01/200319_strain_effect_all//}}
  \centering
\vspace{-1.0cm}
\input{template/change_font_10.tex} 
  \scalebox{0.7}{\input{./data/FLAME_PLANAR_DNS_01/200319_strain_effect_all/planar_dns_jPDF_HR_vs_strain_curv0_TmaxHR_P01_ltx_JFM_HALF.tex}}
\input{template/change_font_12.tex} 
\end{minipage}
\vspace{-1.0cm}
\begin{minipage}[b]{0.45\textwidth}
\input{template/change_font_10.tex} 
\quad(a)
\input{template/change_font_12.tex} 
\end{minipage}
\vspace{0.5cm}
\begin{minipage}[b]{0.45\textwidth}
\input{template/change_font_10.tex} 
\quad(b)
\input{template/change_font_12.tex} 
\end{minipage}
\vspace{0.5cm}
\caption{{\color{black}Planar Flame: Joint-PDFs of the heat release rate and the tangential strain rate for the ${\mathrm{Le}=1}$ dataset~(a) and the engine-relevant ${\mathrm{Le}>1}$ flame~(b). The iso-temperature $T_{\mathrm{maxHR}}$ corresponds to the maximum heat release rate in the corresponding laminar unstretched premixed flames. Data has been conditioned on $\kappa=0$.} }
\label{fig:planar_jPDF_hr_vs_strain}
\end{figure}

{\color{black}The preceding analyses were focussed on the coupling between flame-internal parameters, i.e.\ quantities that respond to changes in heat release rate either due to a non-negligible chemical source term in their transport equation (e.g.\ \mbox{($\left|\nabla c \right|$)-Eq.})~\cite{Kim07_sc_align,Wang17_jet_dns}, or due to a strong heat-release-rate dependence of at least one of their key parameters (cf.\ \mbox{($\mathfrak{D}_{\mathrm{Le}}$)-Eq.}). Here, the flame-external effect of turbulence on the coupled convection/reaction/diffusion system will be discussed. Specifically, the tangential strain rate in the \mbox{($\left|\nabla c \right|$)-Eq.} will be considered as a major entry point of hydrodynamic strain into the flame structure. For the sake of generality, the limit of zero curvature will be investigated in the planar flame datasets, which is consistent with the results presented in Fig.~\ref{fig:planar_h_phi_vs_grad} and enables a phenomenological comparison to strained counterflow flames. It should be noted that in such canonical, steady flame configurations, the mass flux variation through the flame is proportional to the tangential strain rate~\cite{vanOijen02_premixedCntFlow_Le,Kim07_sc_align}. However, it was found that even in steady flames, the parameterization of the burning velocity in terms of the strain rate is non-unique, i.e.\ differs in back-to-back as opposed to fresh-to-burned counterflow flames~\cite{Nilsson02_counterflow,Hawkes06_counterflow}, at least for high strain rates. In turbulent flames, some correlation between the local heat release rate and the tangential strain rate was observed in unity-Lewis-number flames~\cite{Haworth92_dns,Jenkins06_kernel_dns_stretch}. By contrast, DNS results for flames with Lewis numbers significantly different from unity indicate only a very weak correlation~\cite{Savard17_nonLe_flamelet_model,Falkenstein19_kernel_Le_I_cnf}, in particular at high Karlovitz numbers~\cite{Savard15_Le_effects_C7H16_highKa}. However, the role of strain has been mainly discussed based on flame data extracted at one or multiple fixed instants of time in a Eulerian reference frame. In this section, results will be provided that suggest to investigate the local impact of strain at given flame particle locations by means of time series analyses in a flame-attached reference frame, particularly in highly unsteady flames with ${\mathrm{Le}\neq1}$. The following considerations are motivated by the significant global net effect of hydrodynamic strain on the flame structure in the present ${\mathrm{Le}>1}$ dataset that was shown by the low averaged heat release rate at $\kappa=0$ in Fig.~\ref{fig:planar_hr_h_phi_h_vs_curv}\,(a). \par
We begin the forward analysis of turbulence effects on the flame structure from the tangential strain rate~$a_{\mathrm{t}}$ in Fig.~\ref{fig:PrefDiff_schematic}. In absence of a heat-release-rate response to strain that would alter the displacement speed, and in the considered limit of $\kappa=0$, the \mbox{($\left|\nabla c \right|$)-Eq.} suggests net production of~$\left|\nabla c \right|$ by extensive (positive) tangential strain. This is demonstrated in Fig.~\ref{fig:planar_jPDF_grad_vs_strain}\,(a) by a strong positive correlation between~$\left|\nabla T \right|$ and~$a_{\mathrm{t}}$ for the ${\mathrm{Le}=1}$ dataset, which was shown to exhibit only small fluctuations in the local heat release rate~\cite{Falkenstein19_kernel_Le_I_cnf}. Note that the temperature iso-surface corresponding to the maximum temperature gradient magnitude in the laminar unstretched premixed flame was chosen for evaluation, since the slope of~$\left|\nabla T \right|$ in temperature space is rather small at this iso-temperature, which should reduce the scatter due to possible shifts (in temperature space) of the local flame profiles. For ${\mathrm{Le}>1}$, $\left|\nabla T \right|$ is seemingly independent of~$a_{\mathrm{t}}$ as shown in Fig.~\ref{fig:planar_jPDF_grad_vs_strain}\,(c). This very different behavior compared to the ${\mathrm{Le}=1}$ dataset is not surprising, since the initial assumption of a small heat-release-rate response to strain does not hold for the ${\mathrm{Le}>1}$ flame. Note that the flame-external effect of strain is very similar in both flames, which is shown by the marginal strain PDFs provided in Fig.~\mbox{S-6} in the supplementary material. In order to investigate the effect of heat release on the strain dependence of~$\left|\nabla T \right|$, the joint-PDFs evaluated for the iso-temperatures corresponding to the maximum heat release rate in the respective laminar unstretched flames are shown in Fig.~\ref{fig:planar_jPDF_grad_vs_strain}\,(b,d). The comparison of both datasets indicates that in the ${\mathrm{Le}>1}$ flame, production of~$\left|\nabla T \right|$ by positive hydrodynamic strain is overshadowed by the secondary effect of heat release-rate-response to strain, which leads to a weak negative correlation between~$\left|\nabla T \right|$ and~$a_{\mathrm{t}}$. This hypothesis will be analyzed in more detail hereafter by considering the correlation between the local heat release rate and strain.} 
{\color{black}In Part~I of the present study~\cite{Falkenstein19_kernel_Le_I_cnf}, an optimal estimator analysis has shown that in the ${\mathrm{Le}=1}$ limit, the parameterization of the local heat release rate with the external effect of~$a_{\mathrm{t}}$ for a given temperature and flame curvature is as well-suited as using the flame-internal parameter~$Y_{\mathrm{H}}$ at a given temperature. For ${\mathrm{Le}>1}$, the flame-external parameter~$a_{\mathrm{t}}$ (in combination with ($T,\kappa$)) results in a substantially higher irreducible error than any of the tested flame-internal parameters. In the present limit of~$\kappa=0$, $T=T_{\mathrm{maxHR,lam}}$, the correlation between~$\dot{\omega}_T$ and~$a_{\mathrm{t}}$ is indeed weak for ${\mathrm{Le}>1}$, but obviously negative as shown in Fig.~\ref{fig:planar_jPDF_hr_vs_strain}\,(b). This is consistent with results from a thermo-diffusively stable $\mathrm{H}_2$/air flame by Chen and Im~\cite{Chen00_H2_dns}, which were extracted from flame regions with negligible curvature. Phenomenologically, the negative correlation between~$\dot{\omega}_T$ and~$a_{\mathrm{t}}$ will feed back into the \mbox{($\left|\nabla c \right|$)-Eq.} leading to a closed-loop system characterized by multiple timescale ratios, such as the strain Karlovitz number~$\mathrm{Ka}_{\mathrm{s}}=t_{\mathrm{f}}\cdot a_{\mathrm{t}}$~\cite{Candel90_fsd_eq}, and diffusive Damk{\"o}hler numbers for the governing parameters (e.g.\ ($h, \phi, Y_{\mathrm{H}}$)), which could be based on a scalar dissipation rate~\cite{Uranakara17_ign_kernel_3d}, or a relevant diffusion term~\cite{Yoo09_diff_Damk}. Finally, it should be noted that the minor variation in~$\dot{\omega}_T$ shown in  Fig.~\ref{fig:planar_jPDF_hr_vs_strain}\,(a) confirms the initial assumption of small heat-release-rate response for the ${\mathrm{Le}=1}$ flame. \par
To summarize, the cause-effect loop inside the flame structure (cf.\ Fig.~\ref{fig:PrefDiff_schematic}) of a turbulent planar flame with ${\mathrm{Le}>1}$ has been discussed in terms of the tangential strain rate, which is here considered as an external effect. Any change in local strain is expected to propagate through the coupled system of governing flame parameters, which all change based on their own characteristic time scales. It is believed that the weak correlation between the heat release rate and the tangential strain rate in ${\mathrm{Le}\neq1}$ flames is a result of this rather loose coupling, which will require deeper analyses to allow for quantitative conclusions. A possible starting point is a flame particle analysis~\cite{Chaudhuri15_flame_particles,Uranakara17_ign_kernel_3d} that provides access to the time histories of the flame parameter balance equation terms along flame particle trajectories, which is a topic suggested for future work.
}

\section*{Conclusions}
\label{sec:conclusions}
Since experimental SI engine studies have provided evidence that differential diffusion effects in fuel/air mixtures with ${\mathrm{Le}>1}$ lead to an increase in CCV, it is desirable to assess the impact already during engine design. To enable the development of models that capture such fuel effects on CCV, the intention of the present work is to improve the understanding of the complex phenomena inside the flame structure during early flame kernel development. To this end, a DNS database designed to be representative for engine part-load conditions has been analyzed in detail according to a suitable expression for the coupling between the local mixture state and the flame geometry and structure. Flame kernel results have been related to canonical turbulent and laminar flame configurations to enhance generality and increase the relevance of the present findings for future modeling efforts. Specifically, the following conclusions can be drawn: \par
\begin{itemize}
\item The key mathematical expression that characterizes differential diffusion effects under the present engine-relevant {\color{black}(part-load)} conditions is Eq.~(\ref{eq:h_term_D_Le_kappa_k_lam}). 
If non-zero, this term leads to a two-way coupling between the local heat release rate and the flame geometry expressed by curvature, and flame structure expressed by a scalar gradient magnitude. In the limit of ${\mathrm{Le}=1}$, the local mixture state is barely affected by flame geometry and structure since ${\mathfrak{D}_{\mathrm{Le}}=0}$, i.e.\ there is mainly a one-way coupling from the heat release rate to scalar gradients and curvature {\color{black}(also refer to Part~I~\cite{Falkenstein19_kernel_Le_I_cnf})}. The substantially larger number of involved diffusive and hydrodynamic time and length scales may explain the excessive variance in heat release rate associated with the two-way-coupled system~\cite{Falkenstein19_kernel_Le_I_cnf}. Further, the \mbox{($\mathfrak{D}_{\mathrm{Le}}$)-Eq.} suggests two flame parameters for the analysis of differential diffusion effects, $\kappa$ and~${\left|\nabla T \right|/\left|\nabla T \right|_{\mathrm{lam}}}$.
\item It has been shown that the global mean curvature of flame kernels reduces the mixture state parameters ($h, \phi, Y_{\mathrm{H}}$) inside the reaction zone, which lowers the heat release rate and scalar gradients. This effect is most pronounced during the initial phase of flame kernel development and can therefore be effectively compensated by ignition energy supply. After ignition effects have decayed, the effect of mean kernel curvature turned out to be small compared to the overall reduction in heat release rate due to ${\mathrm{Le}>1}$  observed in a statistically planar flame (under the present conditions). However, the effect of global mean curvature is dominant over the impact of the characteristic curvature PDF shape caused by early flame kernel/turbulence interactions~\cite{Falkenstein19_kernel_Le1_jfm}.
\item As suggested by the \mbox{($\mathfrak{D}_{\mathrm{Le}}$)-Eq.}, in case of ${\kappa=0}$ and $\left|\nabla T \right|=\left|\nabla T \right|_{\mathrm{lam}}$, the turbulent flames should locally recover the behavior of an unstretched laminar flame. This has been confirmed by~$\dot{\omega}_T$ and ($h, \phi, Y_{\mathrm{H}}$) dependencies on~$\left|\nabla T\right|$ that are almost identical to the laminar premixed counterflow configuration, {\color{black}which indicates} that differential diffusion effects are not weakened by turbulent micro-mixing {\color{black}under engine part-load conditions. This finding is consistent with previous results by Savard and Blanquart~\cite{Savard17_nonLe_flamelet_model}}.
\item {\color{black}By considering turbulent planar flame elements with zero curvature, hydrodynamic strain was shown to substantially reduce the heat release rate in the ${\mathrm{Le}>1}$ flame on average. However, the local tangential strain rate at a given point in time was found to be only weakly correlated with the local heat release rate for ${\mathrm{Le}>1}$, which is in agreement with previous studies~\cite{Savard17_nonLe_flamelet_model,Savard15_Le_effects_C7H16_highKa}. Here, it is argued that strain alters the two-way coupled system $\left[\left|\nabla c \right|-\left(T, h, \phi, Y_{\mathrm{H}} \right) - \dot{\omega}_c - \left|\nabla c \right|\right]$ mainly through the \mbox{($\left|\nabla c \right|$)-Eq.}, which in turn contains the heat-release-rate response. Since the dynamic propagation of strain effects through the coupled system of governing flame parameters is characterized by several hydrodynamic, diffusive and chemical time scales, it is suggested to investigate the effect of strain on unsteady ${\mathrm{Le}\neq1}$ flames by means of time series analyses in a flame-attached reference frame in the future.}
\item The presented parameter correlations for the flame structure of flame kernels overall recover the behavior of the fully developed turbulent planar flame, after the excess enthalpy and high scalar gradients introduced by spark ignition have decayed within approximately two laminar flame times (under the present conditions). 
\end{itemize}
It should be noted that the overall detrimental behavior of the stoichiometric iso-octane/air flames compared to the ${\mathrm{Le}=1}$ reference flames presented in this work directly translates to common transportation fuels which feature Lewis numbers larger than unity. Although the Reynolds number in the present DNS database is at least two times smaller than in practical engines, the main conclusions should remain valid under more realistic conditions since the Karlovitz number has been closely matched (cf.\ Tab.~\ref{tab:dns_params}). More detailed investigations on the microscopic role of spark ignition during early flame kernel development are suggested for future work.
\section*{Acknowledgement}
The authors from RWTH Aachen University gratefully acknowledge partial funding by Honda R\&D and by the Deutsche Forschungsgemeinschaft (DFG, German Research Foundation) under Research Unit FOR~2687.\par
The authors gratefully acknowledge the Gauss Centre for Supercomputing e.V. (www.gauss-centre.eu) for funding this project by providing computing time on the GCS Supercomputer Super-MUC at Leibniz Supercomputing Centre (LRZ, www.lrz.de).\par
Data analyses were performed with computing resources granted by RWTH Aachen University under project thes0373. \par
S.K. gratefully acknowledges financial support from the National Research Foundation of Korea (NRF) grant by the Korea government (MSIP) (No. 2017R1A2B3008273). \par
T.F. would like to thank G{\"u}nter Paczko,  sadly no longer with us, for the helpful discussions in all these years.
\appendix
\setcounter{figure}{0}
\section{Assessment of a Mean-Curvature-Based Modeling Approach}
\label{apx:model_assess}
\begin{figure}
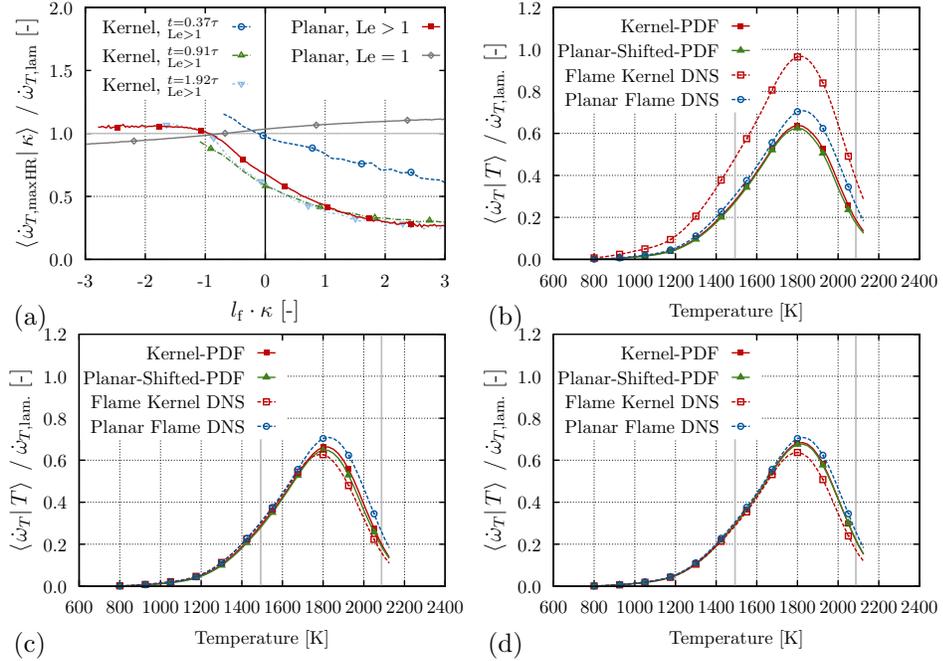
 
\centering
\begin{minipage}[b]{0.45\textwidth}
  \graphicspath{{./data/FLAME_KERNEL_DNS_01/190316_G_local_PHI_curv/}}
  \centering
\input{template/change_font_10.tex} 
  \makebox[0pt][l]{\quad(a)}\scalebox{0.7}{\input{./data/FLAME_KERNEL_DNS_01/190316_G_local_PHI_curv/kernel_dns_CondMean_PDF_HR_CondMaxHR_vs_Curv_K01_LOC0123_P01_P02_allTimes_ltx_JFM_HALF.tex}}
\input{template/change_font_12.tex} 
\end{minipage}
\begin{minipage}[b]{0.45\textwidth}
  \graphicspath{{./data/FLAME_KERNEL_DNS_01/200325_curv_pdf_model_hr/}}
  \centering
\input{template/change_font_10.tex} 
  \makebox[0pt][l]{\quad(b)}\scalebox{0.7}{\input{./data/FLAME_KERNEL_DNS_01/200325_curv_pdf_model_hr/curvPDF_effect_HR_T_LOC01_P01_JFM_HALF.tex}}
\input{template/change_font_12.tex} 
\end{minipage}
\begin{minipage}[b]{0.45\textwidth}
  \graphicspath{{./data/FLAME_KERNEL_DNS_01/200325_curv_pdf_model_hr_t1p5/}}
  \centering
\input{template/change_font_10.tex} 
  \makebox[0pt][l]{\quad(c)}\scalebox{0.7}{\input{./data/FLAME_KERNEL_DNS_01/200325_curv_pdf_model_hr_t1p5/curvPDF_effect_HR_T_LOC01_P01_JFM_HALF.tex}}
\input{template/change_font_12.tex} 
\end{minipage}
\begin{minipage}[b]{0.45\textwidth}
  \graphicspath{{./data/FLAME_KERNEL_DNS_01/200325_curv_pdf_model_hr_t3p3/}}
  \centering
\input{template/change_font_10.tex} 
    \makebox[0pt][l]{\quad(d)}\scalebox{0.7}{\input{./data/FLAME_KERNEL_DNS_01/200325_curv_pdf_model_hr_t3p3/curvPDF_effect_HR_T_LOC01_P01_JFM_HALF.tex}}
\input{template/change_font_12.tex} 
\end{minipage}
\caption{{\color{black}Curvature-conditioned mean heat release rate on the $T_{\mathrm{maxHR}}$-iso-surface~(a). Convolution of two curvature PDFs with the heat release rate conditioned on curvature of a fully developed planar flame at $t=0.37\tau$~(b), at $t=0.91\tau$~(c), and at $t=1.92\tau$~(d). For the flame kernel results, the earlier two times are computed with four realizations, the last time with only one.}}
\label{fig:conv_curv_pdf_hr_vs_T_conv}
\end{figure}

{\color{black}Recall that in Fig.~\ref{fig:hr_curv_pdf_vs_time}, the effect of positive global mean flame kernel curvature on the mean heat release rate of flame kernels has been estimated. This analysis was based on the convolution of a modified, time varying curvature PDF of the developing planar flame with the heat release rate of the planar flame in the fully developed limit. In this way, differences in the curvature distribution between both flame configurations could be isolated from differences in the curvature-conditioned mean heat release rate. Here, this approach is revisited since it can be relevant for modeling. In Fig.~\ref{fig:conv_curv_pdf_hr_vs_T_conv}\,(a), the curvature-conditioned heat release rate during early flame kernel development on the $T_{\mathrm{maxHR}}$-iso-surface is compared against the planar flame results that were presented in Fig.~\ref{fig:planar_hr_h_phi_h_vs_curv}\,(a). Initially, the effect of spark  ignition is dominant, which decays within less than one integral time scale. Note that for $\left|\kappa \right|<l_{\mathrm{f}}$, the heat release rate of the flame kernels then drops slightly below the curve corresponding to the fully developed planar flame that was used for the convolution-based analysis presented in Sect.~\ref{ssec:geom_struct}. This might be due to the fact that in the flame kernel configuration, local flame segments fluctuate around a positive mean curvature with on average lower mean heat release rate than in the planar flame with near-zero mean curvature. If the convolution-based analysis approach pursued in Sect.~\ref{ssec:geom_struct} shall be extended to flame kernel modeling, the observed discrepancies in the curvature-conditioned mean heat release rate of flame kernels and the developed flame will lead to inaccuracies, particularly during the early flame kernel development phase.\par
  In the following, the performance of a model given by the curvature PDF of the fully developed planar flame, but corrected by the mean flame kernel curvature, convoluted with the heat release rate of the fully developed planar flame will be tested. Similar to the analysis presented in Sect.~\ref{ssec:geom_struct}, the convolution has been evaluated for both the modeled and the actual flame kernel PDF, but for multiple temperature iso-surfaces throughout the flame structure. In Fig.~\ref{fig:conv_curv_pdf_hr_vs_T_conv}\,(b,c,d), the results are compared against the actual flame kernel and planar flame DNS data at three different time instants. The similarity of both convolutions confirms the dominant effect of mean flame kernel curvature over the actual shape of the PDF (characterized by higher order moments). After the initial ignition phase, the model predicts a heat release rate that falls between the two DNS results, as expected. In order to develop a more accurate model based on the present approach, a description of the curvature-conditioned heat release rate of flame kernels would need to be incorporated.} 
%
%
%
\FloatBarrier
\bibliography{./references/flame_kernel_01,./references/03_b_Publikationen_anderer.bib,./references/les4ice16.bib,./references/rezchikova.bib}

\end{document}